\documentclass[12 pt]{article}

\usepackage{graphics}
\usepackage{amssymb}
\usepackage[driver]{graphicx}

\newcommand{\N}{\mathbb{N}}
\newcommand{\R}{\mathbb{R}}
\newcommand{\Z}{\mathbb{Z}}

\newcommand{\OO}{{\cal O}}

\newtheorem{claim}{Claim}[section]
\newtheorem{theorem}[claim]{Theorem}
\newtheorem{example}[claim]{Example}
\newtheorem{remark}[claim]{Remark}

\begin{document}
\title{Quantum mechanics of layers with a finite number of
point perturbations}
\date{}
\author{P.~Exner,$^{1,2}$ K.~N\v{e}mcov\'{a}$^{1,3}$}
\maketitle
\begin{quote}
{\small \em 1 Nuclear Physics Institute, Academy of Sciences,
25068 \v Re\v z near \phantom{a)}Prague, Czechia\\ 2 Doppler
Institute, Czech Technical University, B\v rehov{\'a} 7,
\\ \phantom{a)}11519 Prague, Czechia \\ 3 Institute of Theoretical Physics,
FMP, Charles University, \\ \phantom{a)}V Hole\v{s}ovi\v{c}k\'ach
2, 18000 Prague, Czechia}
\end{quote}

\begin{abstract}
\noindent We study spectral and scattering properties of a
spinless quantum particle confined to an infinite planar layer
with hard walls containing a finite number of point perturbations.
A solvable character of the model follows from the explicit form
of the Hamiltonian resolvent obtained by means of Krein's formula.
We prove the existence of bound states, demonstrate their
properties, and find the on-shell scattering operator.
Furthermore, we analyze the situation when the system is put into
a homogeneous magnetic field perpendicular to the layer; in that
case the point interactions generate eigenvalues of a finite
multiplicity in the gaps of the free Hamiltonian essential
spectrum.
\end{abstract}



\section{Introduction}

The object of our interest in this paper is a spinless quantum
particle living in a layer of a fixed width $d$ with the Dirichlet
boundary conditions and interacting with a finite number of point
perturbations. An obvious motivation for this problem is to find a
description for an electron in a semiconductor layer with
impurities. However, such physical systems are in reality rather
complicated objects which involve a crystal lattice with some
alien atoms and an electron gas, so one has to ask first whether
such a simple model can reproduce the basic features known from
experiments.

It is well known that an electron in an ideally pure bulk
semiconductor material can be modeled as a free particle with an
effective mass $m^*$ which characterizes the relation between the
energy and quasi-momentum at the Fermi level. Properties of the
crystalline structure are thus expressed through a single material
constant, which may be very different from the ``bare'' mass --
recall that for GaAs we have $m^*=0.067 m_e$.

There are two other assumptions in the ``free'' part of the model.
The first is its one-particle character which neglects the
interactions between the electrons. There are situations where the
repulsion plays an important role, such as the Coulomb blockade in
quantum wires. On the other hand, the one-electron model is known
to work when the electron-gas density is sufficiently low. Another
assumption is the neglection of spin which is also not entirely
trivial; recall that spin-dependent effects in nanostructures have
been studied recently -- see \cite{BPS} and references therein. In
most situations, however, spinless electrons are a reasonable
approximation.

The next question concerns the way in which we model the
impurities. Using again a certain idealization, we describe them
by point interactions. This method proved rather useful in the
last two decades and gave rise to numerous solvable models; our
aim here is to add one more class to this family. Intuitively
point interactions are understood as sharply localized potentials,
but it is known that a sophisticated coupling constant
renormalization is required to give this concept meaning in terms
of a limit of scaled potentials \cite[Secs.~I.1, I.5]{AGHH}.
Mathematically such operators can be handled since they differ
from the free Hamiltonian $H_0$ just by a change of the boundary
conditions at the interaction sites. However, the counterintuitive
features of three-dimensional point interactions are reflected
both in the slow way in which they found their place in the theory
and in the fact that the parameters appearing in these conditions
cannot be interpreted as $\delta$ potential coupling constants but
rather as the inverse scattering lengths corresponding to the
point ``obstacles'' -- cf.~\cite[Chap.~I.3]{AGHH}.

Scores of papers discussing point-interaction models in the
Euclidean space, both for particles otherwise free or with a
background regular potential, are summarized in the monograph
\cite{AGHH}. Only in the last decade the attention shifted to
systems with point interactions restricted to a certain region of
configuration space; the reason clearly was a wide collection of
new physical phenomena observed in such spatially restricted
systems, mostly mesoscopic objects, but also electromagnetic
waveguides, photonic crystals, etc. -- see \cite{LCM}. Here, too,
point-interaction Hamiltonians proved as a useful tool and yielded
some unexpected results such as the existence of a chaotic
behaviour in systems whose classical counterparts are integrable
\cite{Se}.

Today there are many papers treating point interaction in
restricted areas; a bibliography is given in the introduction of
\cite{EGST}. They typically put emphasis on the description of a
specific model rather than a proper handling of the point
interaction. Among few existing rigorous treatments of the problem
it is the paper \cite{EGST} which motivates the present study
analyzing point interactions in an infinite planar strip with
Dirichlet boundary conditions, together with similar systems.
There are two ways in which the results can be generalized to
dimension three. One is a straight Dirichlet tube in $\R^3$ with a
fixed compact cross section discussed in \cite{E1}; it is a
straightforward extension, apart of a different way of computing
the regularized Green's function.

In the present paper we are going to study a less trivial
generalization, with point interactions situated in a flat layer
with a Dirichlet boundary. The free system allows here again a
separation of variables, so the free resolvent kernel and all the
quantities derived from it such as eigenfunctions, etc., can be
written by means of explicitly given series (in this sense models
considered here are little ``less solvable'' than those in the
full space when such quantities can be written in terms of
elementary or special functions).

Although the model description is simple, it covers many different
situations. For the sake of brevity we restrict ourselves in this
paper to systems with a finite number of point perturbations in
absence of a background potential leaving other cases to a sequel.
We make an exception, however, by devoting a separate section to
the case when the particle is under influence of a homogeneous
magnetic field perpendicular to the layer. The spectrum of the
unperturbed system is then changed completely consisting of
infinitely degenerate eigenvalues which are sums of the Landau
levels and the transverse eigenvalues; for ``rational''
combinations of parameters different Landau levels may lead to the
same eigenvalue. A finite number of point perturbations then gives
rise to a nontrivial discrete spectrum.

Let us describe briefly the contents. The next section is devoted
to the case of a single perturbation. We start from the definition
of the point-interaction Hamiltonian by means of boundary
conditions coupling generalized boundary values. After that we use
Krein's formula to derive the explicit expression for the
resolvent; it involves the regularized Green's function which is
given by a specific series as mentioned above -- see
(\ref{finalxi}). In Section~2.4 we use this result to analyze
spectral properties of such Hamiltonians. The bound state energies
are given by the implicit equation (\ref{point}), and it is just
the limits of strong and weak coupling where we are able to write
the explicit expressions for the leading term of the asymptotics.
In both the extreme cases the eigenvalue behavior can be easily
understood: in the strong-coupling situation it goes to $-\infty$
in the same way as if there were no boundaries because the
corresponding eigenfunction is strongly localized, while in the
weak-coupling case the eigenvalue approaches the threshold of the
essential spectrum and the wavefunction is dominated transversally
by the lowest mode. We also find that the eigenvalue decreases
with the distance from the layer boundary. In the last part of
Section~2 we shall discuss the scattering in the presence of a
perturbation. If there is a single point interaction, we can
employ the symmetry of the problem with respect to rotation around
the axis passing through the perturbation and perpendicular to the
layer. The partial wave decomposition in the ``longitudinal''
coordinates shows that the only nontrivial contribution to the
scattering comes from the s-wave, i.e. from states with the
orbital momentum $m=0$. Within this subspace, the scattering
problem is reduced to transitions between transverse modes; the
final S-matrix then describes a coupling of the ``open'' channels,
i.e. the transverse modes with the energies lower than that of the
incoming spherical wave. We also derive the on-shell scattering
operator which maps the incoming wave vector and transverse mode
into the outgoing ones; the advantage of this approach is that it
does employ the symmetry and allows for a generalization to the
case with multiple perturbations.

Section~3 extends the described analysis to any finite number $N$
of point perturbations. The technique remains the same, and since
the difference between the two resolvents is of rank $N$, the
essential and absolutely continuous spectra are again preserved.
On the other hand, the analysis of the discrete spectrum becomes
more complicated. There are $n$ eigenvalues, where $1\le n\le N$,
which are found by solving the implicit equation $\det
\Lambda(z)=0$ with the $N\times N$ matrix $\Lambda$ given by
(\ref{Lambda}). The number $n$ depends on the coupling strength.
In the strong coupling limit there exist exactly $N$ eigenvalues
having the same asymptotic behavior as in the one-center case. On
the other hand, for weak coupling we find only one eigenvalue
approaching the threshold of the essential spectrum; in this sense
our system exhibits a behavior typical for all weakly coupled
Schr\"odinger operators.

A new feature for systems with $N\ge 2$ is that they can posses
eigenvalues embedded in the essential spectrum. This is possible,
e.g., when the point perturbations are placed symmetrically with
respect to the layer axis and have the same (sufficiently strong)
coupling: the corresponding eigenfunction cannot then contain
contributions from transverse modes with the energy equal or
smaller than this eigenvalue. We will show that this is true for
embedded eigenvalues generally: their eigenvectors have to be
orthogonal to the subspace spanned by the ``lower'' transverse
modes. For $N \ge 2$ the system exhibits no longer a rotational
symmetry, hence we cannot employ the partial-wave decomposition to
describe the scattering. However, the second approach mentioned
above is applicable here and we can derive again the on-shell
scattering operator. It is similar to that of the one-centre case
differing just by replacement of a single regularized Green's
function by a sum of the elements of the matrix $\Lambda$ -- see
(\ref{N on-shell}).

Section~4 deals with the situation when the layer is placed into a
homogeneous magnetic field perpendicular to its boundary. The
Krein's formula is applicable but the free resolvent is
substantially different from the non-magnetic case; this is
reflected in the form of the essential spectrum which now consist
the ``sum'' of the Landau levels and the transverse mode energies;
it, of course, is preserved by a finite number of point
perturbations. If there is a single perturbation we get exactly
one eigenvalue in each spectral gap, i.e. between any two
neighboring levels. In the strong and weak coupling limits it
approaches the upper and lower endpoint of corresponding ``free''
gap, respectively. Only for the lowest gap we find a different
behavior in the strong-coupling limit case; the eigenvalue goes to
$-\infty$ with the same asymptotics as in the non-magnetic case.
Finally we present a generalization to the case of $N$ point
interactions analogous to the considerations of Section~3.


\section{A Single Perturbation}

\subsection{The free system}

Consider an infinite layer $\Sigma := \R^2 \times [0,d]$ with the
coordinates denoted as $\vec x=(x,y)\, $, where $x=(x_1,x_2)\in
\R^2 $ and $y\in [0,d]$. We consider a single spinless
nonrelativistic particle confined to $\Sigma$. Since the actual
values of physical constants are not essential throughout the
paper, we put $\hbar = 2m = 1$ and suppose that the free motion of
the particle is governed by the Dirichlet Laplacian
$-\Delta_D^{\Sigma}$.

Recall that this operator can be defined for rather general
domains in $\R^n$ as the Friedrichs extension of an appropriate
quadratic form \cite[Sec.~XIII.15]{RS}. However, since the
boundary of $\Sigma $ is consists of two disjoint planes and has
therefore the segment property, the operator acts simply as
$H_0\psi = -\Delta\psi $ on the domain of all $\psi$ of the local
Sobolev space $W^{2,2}(\Sigma)$ which satisfy the boundary
conditions
  \begin{equation} \label{layer conditions}
\psi (x,0) = \psi(x,d) = 0
  \end{equation}
for all $x\in\R^2$ -- see again \cite[Sec.~XIII.15]{RS}.

We will make use of the fact that the ``longitudinal" and
``transverse" variables decouple in the free system. The state
Hilbert space of our problem can be then decomposed into traverse
modes, $L^2(\Sigma)=\bigoplus_{n=1}^{\infty} L^2(\R^2) \otimes
\{\chi_n\}$, because the functions $\chi_n(y):=\sqrt{{2 \over
d}}\sin({\pi ny \over d})$ form an orthonormal basis in
$L^2([0,d])$. The free Hamiltonian can be correspondingly written
in the form of a direct sum,
  \begin{equation} \label{decomposition}
 H_0 = \bigoplus_{n=1}^{\infty}\,h_n\otimes I_n  \,, \quad\;
 h_n := -{\partial^2 \over \partial x_1^2}-{\partial^2 \over
 \partial x_2^2}+\Bigl({\pi n \over d}\Bigr)^2.
  \end{equation}
Since the resolvent $(H_0-z)^{-1}$ of the two-dimensional
Laplacian is known explicitly, the above decomposition yields in
turn the free resolvent kernel
  \begin{equation} \label{free resolvent}
 G_0(x,y;x',y';z) = {i \over 4}\sum_{n=1}^{\infty}
 H_0^{(1)}(k_n|x-x'|)\, \chi_n(y)\,\chi_n(y'),
  \end{equation}
where $k_n\equiv k_n(z) :=\sqrt{z-({\pi n \over d})^2}\, $.

\subsection{Definition of a point interaction}

Our first goal is to construct a one-center perturbation supported
by a point $\vec a :=(a,b)$ with $a\in\R^2$ and $b\in(0,d)$. This
can be done in a standard way \cite{AGHH}. We restrict
$-\Delta_D^{\Sigma}$ to functions which vanish in a neighborhood
of $\vec a$; the operator obtained in this way is symmetric but
not self-adjoint and we look for the perturbed one among its
self-adjoint extensions. Since the restriction has deficiency
indices $(1,1)$, the family of extensions is by the standard von
Neumann theory \cite[Sec.~X.1]{RS} characterized by a single
parameter.

What is equally important is that the perturbation is local, and
therefore we can characterize the extensions by the usual boundary
condition derived in \cite[Chap.~I.1]{AGHH} for point interactions
in $\R^3$. We introduce the generalized boundary values,
$$L_0(\psi,\vec a) := \lim_{|\vec x -\vec a| \to 0}|\vec x -\vec
a|\,\psi(\vec x), \,\quad L_1(\psi,\vec a) := \lim_{|\vec x -\vec
a|\to 0} \biggl\lbrack\psi(\vec x)-{L_0(\psi,\vec a) \over |\vec
x-\vec a|}\biggr\rbrack\,,$$
and require
  \begin{equation} \label{bc}
L_1(\psi,\vec a) -4\pi\alpha L_0(\psi,\vec a) = 0\,.
  \end{equation}
For a fixed $\alpha\in\R$ this leads to the self-adjoint operator
$H(\alpha,\vec a)$ acting as
  \begin{equation} \label{one-center perturbation}
(H(\alpha,\vec a)\psi)(\vec x) = -(\Delta\psi)(\vec x)
  \end{equation}
for $\vec x\ne\vec a$ on the domain
  \begin{equation} \label{one-center domain}
 D(H(\alpha,\vec a)) := \Bigl\lbrace\psi\, :\,
 -\Delta\psi\in L^2 (\Sigma \setminus\{\vec a\}) {\rm \; and\;
 (\ref{layer conditions}),\; (\ref{bc})\; are\; satisfied}
 \Bigr\rbrace
  \end{equation}
in $L^2(\Sigma)$, where $-\Delta\psi$ is, of course, understood in
the sense of distributions. The family of self-adjoint extension
includes also case which is formally given by $\alpha =\infty$,
which means $L_0(\psi,\vec a)=0$. It is easy to see that the
corresponding $H(\alpha,\vec a)$ is nothing else than the free
Hamiltonian $H_0$.

\subsection{The resolvent} \label{resol 1}

As usual the spectral properties of the operator $H(\alpha,\vec
a)$ can be studied using its resolvent. Since $H(\alpha,\vec a)$
and $H_0$ have a common restriction with deficiency indices
$(1,1)$, the kernel of the full resolvent can be obtained by means
of the Krein's formula \cite[Appendix A]{AGHH}
  \begin{equation} \label{krein}
(H(\alpha,\vec a)-z)^{-1}(\vec x_1,\vec x_2) = G_0(\vec x_1,\vec
x_2 ;z)\,+\,{G_0(\vec x_1,\vec a ;z)\, G_0(\vec a,\vec x_2 ;z)
\over \alpha-\xi(\vec a;z)}\,,
  \end{equation}
where
  \begin{equation}\label{ksi0}
\xi(\vec a;z) := {1 \over 4\pi}\lim_{\vec x\to \vec a}\biggl
\lbrace {2\pi i \over d}\sum_{n=1}^{\infty}H_0^{(1)}(k_n|x-a|)\sin
\Bigl({\pi ny \over d}\Bigr)\sin\Bigl({\pi nb \over d}\Bigr)-{1
\over |x-a|}\biggr\rbrace
  \end{equation}
is the regularized Green's function; we have employed here the
fact that the resolvent singularity is the same as for the kernel
of $-\Delta$ in $L^2( \R^3)$ -- see \cite[Sec.~13.5]{Titch}.
The form of the denominator in expression (\ref{krein}) follows from
the boundary condition (\ref{bc}) applied to $\psi=(H(\alpha,\vec a)
-z)^{-1} \varphi$, where $\varphi$ is an arbitrary vector from $L^2
(\Sigma)$.
However, the above definition does not give a practical way to
compute $\xi(\vec a;z)$. To this end we use first $K_0(z)={\pi i
\over 2} H_0^{(1)}(iz)$ and introduce $\kappa_n:=\sqrt{({\pi n
\over d})^2-z}=-i k_n$, then we have
  \begin{equation}\label{ksi}
  \xi(\vec a;z) = \lim_{\varrho\to 0}\biggl\lbrace {1 \over \pi d}
 \sum_{n=1}^{\infty} K_0(\kappa_n\varrho) \sin^2 \Bigl({\pi nb
 \over d} \Bigr)\,-\,{1 \over 4\pi\varrho} \biggr\rbrace\,,
  \end{equation}
where $\varrho:=|x-a|$ and we have already put $y=b$. We use the
asymptotic behavior $\kappa_n\approx {\pi n \over d}$ as
$n\to\infty$ to write the identity
  \begin{equation}
 K_0(\kappa_n\varrho) = K_0\Bigl(n {\pi\varrho \over d}\Bigr)
 +\Bigl\lbrack K_0(\kappa_n\varrho)-K_0\Bigl(n {\pi\varrho \over d}
 \Bigr)\Bigr\rbrack\,.
  \end{equation}
and to divide the function $\xi$ into two parts, $\xi(\vec
a;z)=\xi_1+\xi_2$, where
  \begin{eqnarray*}
\xi_1 &\!:=\!& \lim_{\varrho\to 0} {1 \over \pi
d}\sum_{n=1}^{\infty}\, \Bigl\lbrack
K_0(\kappa_n\varrho)-K_0\Bigl( {\pi n\varrho \over d}
\Bigr)\Bigr\rbrack\sin^2\Bigl({\pi nb \over d}\Bigr)\,,
\\ \xi_2 &\!:=\!& \lim_{\varrho\to 0} \biggl\lbrace {1 \over 2\pi
d}\sum_{n=1}^{\infty}\, \Bigl\lbrack K_0\Bigl({\pi n\varrho \over
d}\Bigr)-K_0\Bigl({\pi n\varrho \over d}\Bigr)\cos\Bigl({2\pi nb
\over d}\Bigr) \Bigr\rbrack -{1 \over 4\pi\varrho}
\biggr\rbrace\,;
  \end{eqnarray*}
we have used $2\sin^2\alpha = 1-\cos 2\alpha$. To deal with the
first term we employ the asymptotic behavior of the Macdonald
function \cite[9.6.13]{AS} which yields
$$K_0(\kappa_n\varrho)-K_0\Bigl({\pi n\varrho \over d}\Bigr) =
-\ln\sqrt{1-z\Bigl({d \over \pi n}\Bigr)^2}\,
\bigl(1+\OO(\varrho^2)\bigr) $$
as $\varrho\to 0$. It shows that the sum converges uniformly
w.r.t. $\varrho$ and the limit can be interchanged with the series
giving thus
  \begin{equation}
\xi_1 = -{1 \over \pi d}\sum_{n=1}^{\infty}\, \ln\sqrt{1-z\Bigl(
{d \over \pi n}\Bigr)^2}\,\sin^2\Bigl({\pi nb \over d}\Bigr)\,.
  \end{equation}
The second part can be computed by means of the formula \cite[II,
5.9.1.4.]{BMP},
  \begin{eqnarray*}
\lefteqn{ \sum_{n=1}^{\infty}\, K_0(nx)\cos(na) = {\pi \over
2\sqrt{x^2+a^2}} + {1 \over 2} \Bigl(\gamma+\ln {x \over 4\pi}
\Bigr) } \\ && \hspace{-7mm} +\, {\pi \over
2}\sum_{n=1}^{\infty}\, \biggl\lbrack {1 \over
\sqrt{(2n\pi-a)^2+x^2}}-{1 \over 2n\pi}\biggr\rbrack +\, {\pi
\over 2}\sum_{n=1}^{\infty}\, \biggl\lbrack {1 \over
\sqrt{(2n\pi+a)^2+x^2}}-{1 \over 2n\pi}\biggr\rbrack,
  \end{eqnarray*}
so introducing $\beta:={b \over d},\; \mu:={\varrho \over d}$ and
performing the limit $\mu\to 0$ , we get
  \begin{equation}
\xi_2 = \,{1 \over 4\pi d}\biggl\lbrace-{1 \over 2\beta}
-\sum_{n=1}^{\infty}{\beta^2 \over n(n^2-\beta^2)}
\biggr\rbrace\,.
  \end{equation}
The last sum equals using \cite[I, 5.1.15.2.]{BMP} expressed by
means of the digamma function as $\psi(1)-{1 \over
2}\bigl(\psi(1\!+\!\beta)+\psi(1\!-\!\beta)\bigr)$, and since
$\psi(1)=-\gamma$, where $\gamma=0.577...$ is the Euler number,
and $\psi(1-\beta)=\psi(\beta)+\pi\cot(\pi\beta),\;
\psi(1+\beta)=\psi(\beta)+{1 \over \beta}$, we arrive at
  \begin{equation}\label{xi2}
\xi_2 = \frac{\gamma}{4\pi d}+{1 \over 8\pi d}\,\Bigl(
2\psi(\beta)+\pi\cot(\pi\beta)\Bigr)\,.
  \end{equation}
Putting the results together, we get the sought formula
  \begin{equation}\label{finalxi}
\xi(\vec a;z) = -{1 \over \pi d}\sum_{n=1}^{\infty}\ln
\sqrt{1\!-\!z\Bigl({d \over \pi n}\Bigr)^2} \sin^2 \Bigl({\pi nb
\over d}\Bigr)+{1 \over 4\pi d}\Bigl\lbrack\gamma+\psi\Bigl({b
\over d}\Bigr) +{\pi\over 2}\cot\Bigl({\pi b \over d}\Bigr)
\Bigr\rbrack
  \end{equation}
expressing the regularized Green's function in the form of a
series. It is certainly more complicated than an expression of the
corresponding quantity for the whole space in terms of elementary
functions \cite[Chap.~I.1]{AGHH}, but it allows us to derive the
needed properties of the function $\xi$ to compute the values of
$\xi(\vec a;z)$ numerically.

  \begin{remark} \label{scaling}
{\rm Notice the scaling behavior with respect to the change of the
layer thickness, i.e. the formulae relating properties of the
family $\Sigma^{\sigma}=\R^2 \times [0, d\sigma], \; \sigma>0$.
Here the dimension of the configuration space is decisive. While
for two-dimensional system the scaling amounts to logarithmic
shift of the function $\xi$ as shown in \cite{EGST}, in three
dimension the transformation is multiplicative. We find easily
that the situation is the same as for straight tubes in $\R^3$
studied in \cite{E1}, i.e. we have $\xi(\vec{a^{\sigma}};
z\sigma^{-2})= \sigma^{-1} \xi(\vec a;z)$, where
$\vec{a^{\sigma}}:= (a\sigma,b\sigma)$. This means, in particular,
that the singularities of the resolvent which we will discuss
below using equation (\ref{point}) are related as follows:
  \begin{equation} \label{scale}
\epsilon^{\sigma}(\alpha^{\sigma},\vec{a^{\sigma}})  = \sigma^{-2}
\epsilon(\alpha,\vec{a})\,, \quad \alpha^{\sigma}:= \sigma^{-1}
\alpha\,.
  \end{equation}
Without loss of generality we put therefore $d=\pi$ in the rest of
this and the next section.}
  \end{remark}

\subsection{Spectral properties} \label{spec 1}

The explicit form of the resolvent (\ref{krein}) allows us to
derive information about the spectrum. Since its difference from
$(H_0-z)^{-1}$ is a rank one operator, the essential spectrum
remains by Weyl's theorem \cite[Thm.~XIII.14]{RS} the same as for
the free Hamiltonian $H_{0}$, i.e., we have
$\sigma_{ess}(H(\alpha, \vec a))=[1,\infty)$. At the same time,
also the absolutely continuous spectrum is preserved,
$\sigma_{ac}(H(\alpha, \vec a))=[1,\infty)$, this time by
Birman-Kuroda theorem \cite[Thm.~XI.9]{RS}.

Next we would like to prove the absence of the singularly
continuous spectrum. To this aim we have to check that the
expression $(\psi, (E_z-E_c) \psi)$ for all $\psi \in L^2(\Sigma)$
is an absolutely continuous function for $z$ from any interval
$(c,t) \subset (1, \infty)$, in other words, that it can be
written as an integral of a locally integrable function. Since
$\sigma_{sc}$ cannot be supported by discrete points, we may
choose the interval $(c,t)$ in such a way that it contains none of
the points $m^2, \; m \in \N$. For the spectral projection $E_t
-E_c$ to the interval $(c,t)$ we employ the Stone's formula,
  \begin{equation}
(\psi, (E_t-E_c) \psi) = {1 \over \pi}\, \lim_{\varepsilon \to
0_+} \int_c ^t \left( \psi, {\rm Im\,}G (u+i\varepsilon) \psi
\right) \, du\,; \nonumber
  \end{equation}
we have used here the fact which we will establish a little later,
namely that $H(\alpha, \vec a)$ has no eigenvalues in
$(1,\infty)$, and therefore the spectral projections to $(c,t)$
and $[c,t]$ are the same. The Green's function (\ref{krein}) is
analytic for $z$ with ${\rm Re\,}z\in(c,t)$ and $\pm {\rm Im\,}z
\in (0,\varepsilon)$, and furthermore, its limits when $z$
approaches the real axis from above and from below exist and are
continuous functions of $z$. Recall that by assumptions no
thresholds are contained in $(c,t)$ and $\xi$ has nonzero
imaginary part for $z\neq m^2$,
  \begin{equation} \label{im part}
{\rm Im}\,\xi(\vec a; z) =  {1 \over 2\pi}\, \sum
\limits_{n=1}^{[\sqrt{z}]}\, \sin^2 (n b) > 0\,,
  \end{equation}
where we have used $\sqrt{z-m^2}=i\sqrt{m^2-z}$; hence the
denominator of the second term in (\ref{krein}) cannot be
singular. Consequently, the integrated function is bounded in
$(c,t)\times[0,\varepsilon)$ and by the dominated convergence
theorem the limits can be interchanged with the integral giving
  \begin{equation}
(\psi, (E_t-E_c) \psi) = {1 \over \pi}\,  \int_c^t \left( \psi,
{\rm Im\,}G (u) \psi \right) \, du.
  \end{equation}
The function under the integral is again continuous in the
interval $(c,t)$, hence it is integrable and the statement is
proved. If $t=t_0$ would be an isolated eigenvalue embedded in the
continuous spectrum and $\psi$ the corresponding eigenfunction,
the above relation remains valid for $t\in(t_0\!-\!\eta,t_0) \cup
(t_0,t_0\!+\!\eta)$ with some $\eta>0$, while at the point $t_0$
the l.h.s. should have a jump, which is clearly impossible due to
the continuity of the integrated function.

To determine the discrete spectrum, we have to find the poles of
the resolvent. Recall that a perturbation which can be reduced to
a self-adjoint extension of a common symmetric restriction with
deficiency indices $(1,1)$ can give rise to at most one simple
eigenvalue in each gap of the spectrum \cite[Sec.~8.3,
Cor.~1]{Wei}. In our case it means one simple eigenvalue in the
interval $(-\infty,1)$. In view of the relation (\ref{krein}) one
can find it by solving the implicit equation
  \begin{equation}\label{point}
\xi(\vec a;z) = \alpha
  \end{equation}
for $z\in\R$. The series contained in the formula (\ref{finalxi})
converges for all $z\in\R \setminus\{n^2: n\in\N\}$, because its
terms decay like $n^{-2}$ as $n \to \infty$ as we can see using
the Taylor expansion of $\ln(1-\zeta)$ to the first order. The
remaining term $\xi_2$ is independent of $z$ and finite for any
$b\in (0,\pi)$.

\begin{figure}[!t]
\begin{center}
\includegraphics[height=10cm, width=\textwidth]{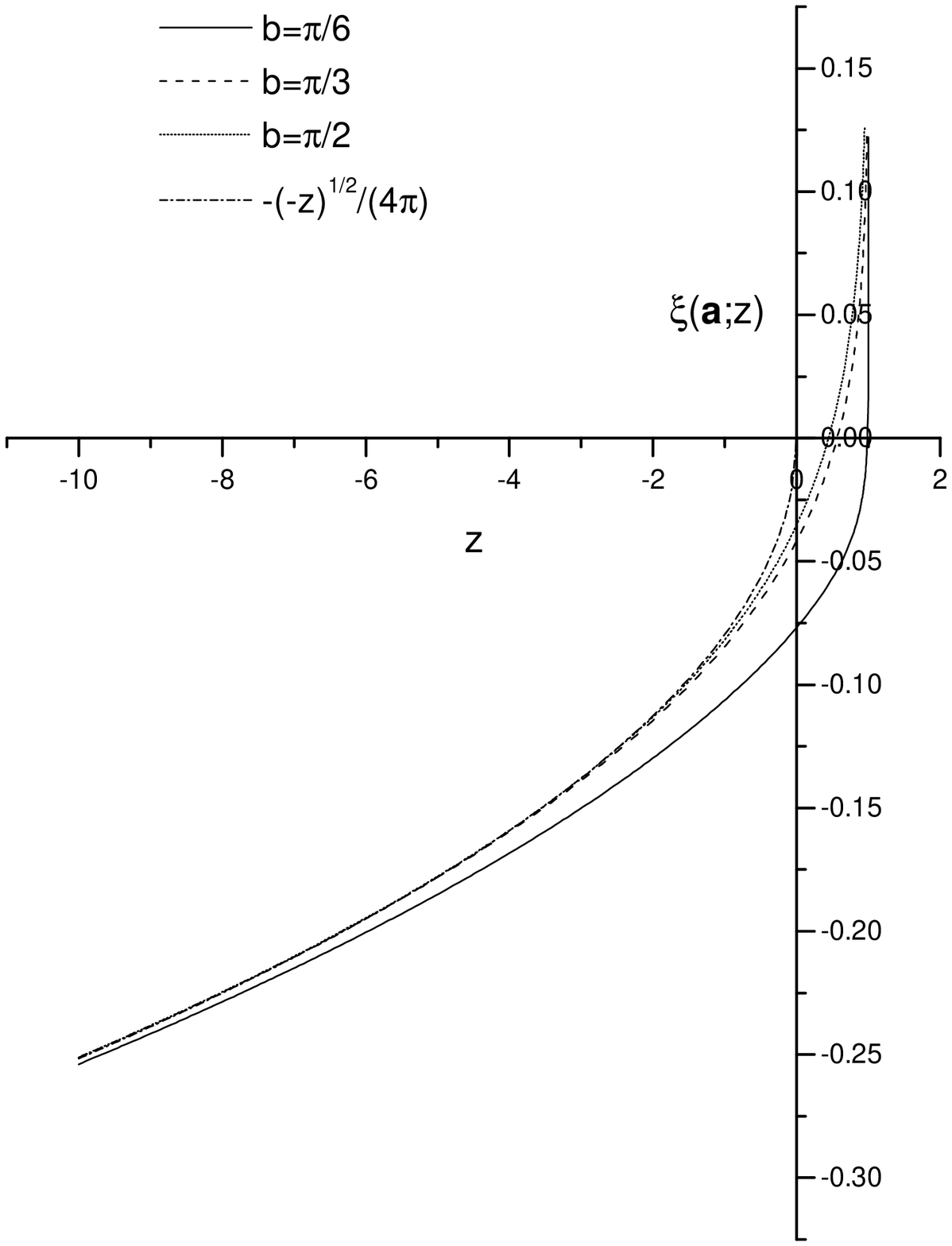}
\end{center}
\vspace{-2cm} \caption{The function $\xi(\vec a; \cdot)$ for three
different positions of the point interaction. The dash-dotted line
is the leading term of the asymptotics (\ref{strong c asympt}).}
\label{fig:xi as}
\end{figure}

The value of $\xi(\vec a;z)$ is real for any $z\in (-\infty,1)$.
In particular, it is easy to compute
$$\xi(\vec a;z=0) = {1 \over 4\pi^2 } \left[ \gamma +\psi \Big( {b
\over \pi} \Big) + {\pi \over 2} \cot (b) \right]\,. $$
Differentiating $\xi(\vec a,\cdot)$ we get
  \begin{equation} \label{dxi}
{d \xi \over dz} = {1 \over 2 \pi^2} \sum \limits_{n=1}^{\infty}\,
{1 \over n^2 - z } \; \sin^2 (n b)\,>\,0\,,
  \end{equation}
so the function is monotonously increasing for $z<1$. Moreover, it
diverges at both endpoints. At the continuum threshold we have
  \begin{equation} \label{threshold asympt}
\xi(\vec a;z) =  -{1 \over \pi^2} \ln \sqrt{1-z}\, \sin^2(b)+
\OO(1) \qquad {\rm as} \qquad z\to 1-\,,
  \end{equation}
while on the opposite side we may employ a simple estimate
$$\xi(\vec a;z)\,<\,-{1 \over \pi^2}\ln \sqrt{1-z}\, \sin^2(b)
+\xi_2\,\to\ -\infty \qquad {\rm as} \qquad z \to -\infty\,.$$
We will need a more precise asymptotics at large negative
energies. Below we shall prove that
  \begin{equation} \label{strong c asympt}
\xi(\vec a;z) = -{\sqrt{-z} \over 4\pi}+
\OO\left(e^{-c\sqrt{-z}}\right) \qquad {\rm as} \qquad z \to
-\infty
  \end{equation}
for any $c<\mathrm{dist}(\vec a,\mathrm{bd\,}\Omega) = {\pi\over
2}-\left|b-{\pi\over 2}\right|$. In other words, the leading term
corresponds to the analogous function for the point-interaction
Laplacian in $ \R^3$ computed in \cite[Chap.~I.1]{AGHH}; it
corresponds to the heuristic concept that strongly bound states
are well localized, and therefore not much influenced by the
presence of the boundary. Another property of $\xi(\vec a,z)$ is
its monotonicity across the halflayer,
  \begin{equation}
\xi (\vec a;z) > \xi (\vec a';z) \qquad {\rm if} \qquad |b-\pi/2|
 < |b'-\pi/2|, \; a=a'. \nonumber
  \end{equation}
To prove it we employ the relation (\ref{ksi0}) which yields
  $$
\xi (\vec a;z) - \xi (\vec a';z) = \lim_{|x- a| \to 0}\: {i \over
2\pi} \sum_{n=1}^\infty H_0^{(1)} (k_n(z) |x-a|) [ \sin^2 (n b) -
\sin^2 (n b') ]\, .
  $$
Since $\sin^2 (n b) - \sin^2 (n b') = \sin (n(b+b')) \sin
(n(b-b'))$ holds for $0<b'<b \leq \pi /2$ we arrive at
  \begin{equation}
\xi (\vec a;z) - \xi (\vec a';z) = G_0 (a,b+b'; a,b-b'; z)\,;
  \end{equation}
the monotonicity then follows from the positivity of free
resolvent kernel -- cf.~\cite[Appendix to Sec.~XIII.12]{RS}. This
behavior is illustrated in Fig.~\ref{fig:xi as}.

\begin{figure}[!t]
\begin{center}
\includegraphics[height=10cm, width=\textwidth]{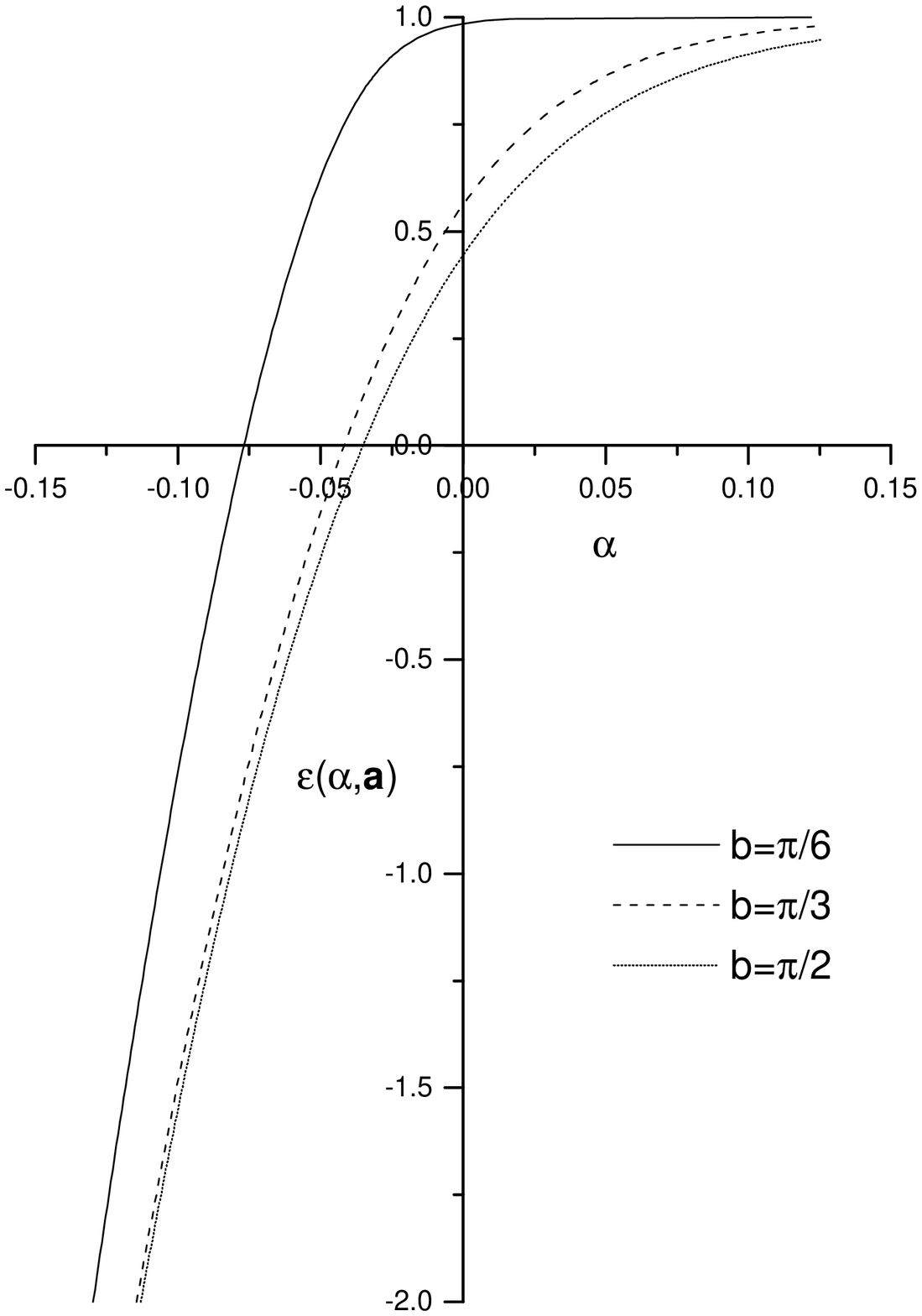}
\end{center}
\vspace{-2cm} \caption{The dependence of the eigenvalues
$\varepsilon(\alpha,\vec a)$ on the parameter $\alpha$ for
 three positions of the point interaction.} \label{fig:inv}
\end{figure}

This confirms the mentioned general conclusion: it follows from
the stated properties of $\xi(\vec a;\cdot)$ that the equation
(\ref{point}) has for any $\alpha\in \R$ a unique eigenvalue
$\varepsilon(\alpha,\vec a)$ in $(-\infty,1]$ and that the
function $\varepsilon(\cdot,\vec a)$ is monotonously increasing,
$$\varepsilon(\alpha,\vec a)>\varepsilon(\alpha',\vec a) \qquad
{\rm if} \qquad \alpha>\alpha'.$$
Furthermore, $\xi(\vec a,\cdot)$ is a real-analytic function
because for a fixed $z_1<-1$ it is expressed on a complex
neighborhood of $(-\infty,z_1)$ through a uniformly convergent
series whose terms are analytic. It follows from the
implicit-function theorem that $\varepsilon(\cdot,\vec a)$ is a
$C^{\infty}$ function -- see \cite[Chap.XIV]{D1}. The function
$\varepsilon(\cdot,\vec a)$ is also monotonous with $b$
  \begin{equation}\label{strip monotonicity}
\varepsilon(\cdot,\vec a) < \varepsilon(\cdot,\vec a') \qquad {\rm
if} \qquad |b- \pi /2| < |b'- \pi /2|\,.
  \end{equation}
The behavior of the eigenvalue is shown in Fig.~\ref{fig:inv}.

We are also interested in the asymptotic behavior of the
eigenvalue in the limits of weak and strong coupling. In the
former case we have
  \begin{equation} \label{weak 1 ev}
\varepsilon(\alpha,\vec a) \approx 1 - \exp \left( - {2 \pi^2
\alpha \over \sin^2 (b)} \right)\qquad {\rm as} \qquad \alpha
\to\infty\,,
  \end{equation}
where the symbol $\approx$ means that for any $\epsilon>0$ and
$\alpha$ large enough the eigenvalue can be squeezed between a
pair of expressions from the r.h.s. in which $2\pi^2$ is replaced
by $2\pi^2\pm\epsilon$. On the other hand, the strong coupling
asymptotics can be proved directly. By Dirichlet bracketing
\cite[Sec.~XIII.15]{RS} $\varepsilon(\alpha,\vec a)$ is for
$\alpha<0$ bounded from below by $-(-4\pi\alpha)^2$, and from
above by the ground state of the Dirichlet Laplacian in a ball of
radius $c$ with the point interaction in the center. The latter is
easily found: writing $\varepsilon(\alpha,\vec a)=-\kappa^2$ one
has to solve the equation
  $$
(-4\pi\alpha)^2 = \kappa^2 (1+\sinh^{-2} \kappa c)\,.
  $$
It yields the sought asymptotic behavior
  \begin{equation} \label{strong 1 ev}
\varepsilon(\alpha,\vec a) = -16\pi^2\alpha^2 +\OO\left(e^{\alpha
c} \right) \qquad {\rm as} \qquad \alpha \to -\infty\,,
  \end{equation}
which justifies in view of (\ref{point}) a posteriori the relation
(\ref{strong c asympt}).

\begin{figure}[!b]
\begin{center}
\vspace{1.5cm} \hspace{5cm}
\includegraphics[height=8cm, width=7cm]{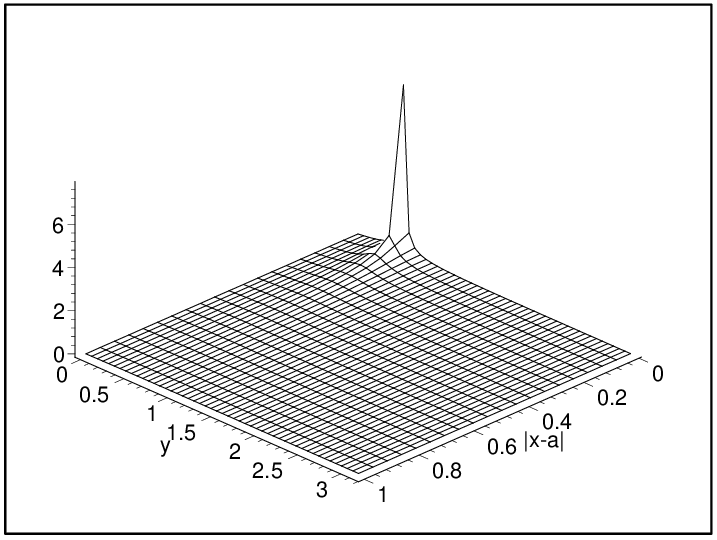}
\end{center}
\vspace{-4cm} \caption{The unnormalized eigenfunction for
$b=\pi/6$ and $\alpha=-0.1$.} \label{fig:ef6}
\end{figure}

The formula (\ref{krein}) provides us with the (non-normalized)
wavefunction of the bound state through the residue at the pole,
$\psi(\vec x; \alpha, \vec a)=G_0(\vec x, \vec a; \varepsilon
(\alpha, \vec a))$, so we have
  \begin{equation} \label{ef1}
\psi(\vec x; \alpha, \vec a) = {i \over 2\pi}\sum_{n=1}^{\infty}
\, H_0^{(1)} \left( \sqrt{\varepsilon (\alpha, \vec a)-n^2}\,
|x-a| \right) \sin(nb)\sin(ny).
  \end{equation}
For $\alpha \to -\infty$ we can write $H_0^{(1)}(u) \approx
\sqrt{{2 \over \pi u}}\,e^{i(u-\pi/4)}$ so we see that the
wavefunction is well localized,
  \begin{equation}\label{strong 1 ef}
\psi(\vec x; \alpha, \vec a) \approx \sum_{n=1}^{\infty} \sqrt{{1
\over \pi^3\sqrt{16\pi^2\alpha^2 +n^2}\,|x-a|}}\:
e^{-\sqrt{16\pi^2\alpha^2+n^2}\,|x-a|}\,\sin(n b) \sin( n y)\,;
  \end{equation}
this is illustrated in Fig.~\ref{fig:ef6}.
On the other hand, in the limit $\alpha \to \infty $ the
wavefunction for $|x-a|$ from a compact set behaves as
  \begin{eqnarray} \label{weak 1 ef}
\psi(\vec x; \alpha, \vec a) &\!\approx\!& \alpha\, {\sin y \over
\sin b}\, -\, {1 \over \pi^2} \ln|x-a|\sin b \sin y  \\ && +\, {i
\over 2\pi}\, \sum_{n=2}^{\infty}\, H_0^{(1)} \left(
\sqrt{1-n^2}\, |x-a|\right) \sin(n b) \sin(n y)\,. \nonumber
  \end{eqnarray}
\begin{figure}[!b]
\begin{center}
\vspace{1.5cm} \hspace{5cm}
\includegraphics[height=8cm, width=7cm]{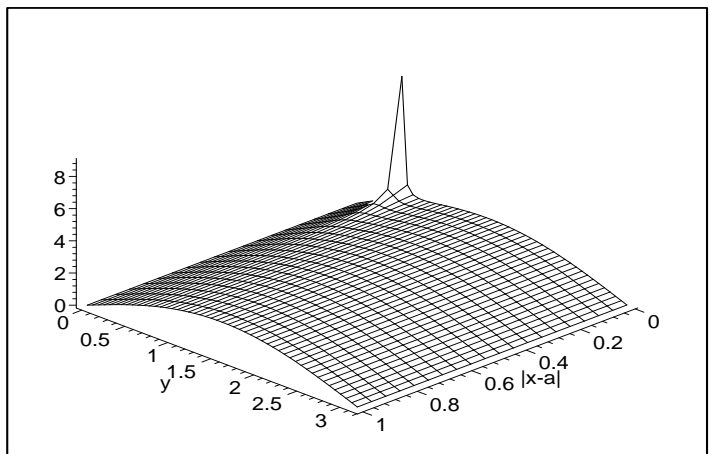}
\end{center}
\vspace{-4cm} \caption{The unnormalized eigenfunction for
$b=\pi/6$ and $\alpha=1$.} \label{fig:ef6 w}
\end{figure}
The wavefunction is dominated by the first transverse mode as
Fig.~\ref{fig:ef6 w} shows.

Let us finally return to embedded eigenvalues. We have excluded
their existence away of the thresholds. If $z=m^2$ we use the
equation (\ref{krein}) again to check that the singularities at
the r.h.s. cancel. In the vicinity of $z=m^2$ the free resolvent
kernel and the denominator $\alpha -\xi(\vec a;z)$ behave as
  \begin{eqnarray*}
 G_0(\vec x_1,\vec x_2;z) &\!=\!& -\,{1 \over
 \pi^2}\ln\sqrt{m^2-z}\, \sin(my_1)\sin(my_2)
 \left(1+\OO\left(\sqrt{m^2-z} \right) \right), \\
 \alpha-\xi(\vec a;z) &\!=\!& \tilde{\alpha}+ {1 \over \pi^2}
 \ln\sqrt{m^2-z}\sin^2(mb)\,,
  \end{eqnarray*}
where
$$\tilde{\alpha} := \alpha-\xi_2+{1\over \pi^2} \sum_{n\neq m}\,
\ln\sqrt{1-{m^2 \over n^2}}\,\sin^2(nb)- {1\over \pi^2}\ln
m\sin^2(mb)\,. $$
Then the full-resolvent kernel asymptotically
behaves as
  \begin{eqnarray*}
\lefteqn{ \left(\tilde{\alpha}+{1 \over
\pi^2}\ln\sqrt{m^2-z}\,\sin^2(mb) \right)^{-1}
{\tilde{\alpha}\over \pi^2} \ln\sqrt{m^2-z}\,
\sin(my_1)\sin(my_2)} \\ && \phantom{AAAAAAAAAAAAAAAAAA} \times
\left(1+\OO\left(\sqrt{m^2-z} \right) \right)
  \end{eqnarray*}
and cannot thus have a pole-type singularity at $z=m^2$.

Let us summarize the results obtained so far:

\begin{theorem}
 Let $H(\alpha, \vec a)$ be defined by (\ref{one-center perturbation})
 and (\ref{one-center domain}) for $d=\pi$; then \\ [1mm]
 (a) $\sigma_{ess}(H(\alpha, \vec a)) = \sigma_{ac}(H(\alpha, \vec a))
 =[1,\infty)$ and $\sigma_{sc}(H(\alpha, \vec a))=\emptyset$.\\ [1mm]
 (b) For any $\alpha\in\R$ there is a single eigenvalue
 $\varepsilon(\alpha,\vec a)$ in $(-\infty,1)$ which is increasing and
 infinitely differentiable w.r.t. $\alpha$. The corresponding
 eigenfunction is given by (\ref{ef1}). \\ [1mm]
 (c) The eigenvalue is by (\ref{strip monotonicity}) strictly monotonous
 across the halflayer, $\varepsilon(\alpha,\vec a) <
 \varepsilon(\alpha,\vec a')\,$ if $\,|b-\pi /2| < |b'-\pi /2|$. \\ [1mm]
 (d) In the limit $\alpha\to\infty$ the bound state behaves
 according to (\ref{weak 1 ev}) and (\ref{weak 1 ef}). In the
 strong coupling case the eigenvalue asymptotics is given by
 (\ref{strong 1 ev}) and the eigenfunction is described by
 (\ref{strong 1 ef}). \\ [1mm]
 (e) There are no eigenvalues in $[1,\infty)$.
\end{theorem}

\subsection{Scattering}\label{scattering 1}

Since the Hamiltonian $H(\alpha, \vec a)$ is invariant under
rotations around the axis passing through the point $\vec a$ and
perpendicular to $\Sigma$, we may simplify the treatment of
stationary scattering using a partial-wave decomposition. We use
the tensor-product representation
  \begin{equation}\label{tensor Hilb}
L^2(\Sigma) = L^2((0,\infty)\times[0,d]; r\,dr\,dy) \otimes
L^2(S^1)\,, \nonumber
  \end{equation}
where $S^1$ is the unit circle in $\R^2$ and $r:=|x|$. This can be
written as
  \begin{equation}\label{tensor Hilb2}
L^2(\Sigma) = \bigoplus_{m=-\infty}^{\infty} \tilde U^{-1}
L^2((0,\infty)\times[0,d]) \otimes \{Y_m\}\,,
  \end{equation}
where $\tilde U:\: L^2((0,\infty)\times[0,d]; r\,dr\,dy) \to
L^2((0,\infty)\times[0,d])$ is the unitary operator defined by
$(\tilde U\psi)(r) := r^{1/2} \psi(r)$, the ``spherical''
functions $Y_m(\omega) := (2\pi)^{-1/2} e^{i m\theta}$ with
$\omega = (\cos\theta, \sin\theta)$ form a basis in $L^2(S^1)$ and
the symbol $\{\cdot\}$ means as above the linear envelope.
Shifting the point $\vec a$ to the origin of the polar coordinates
by $T_a:\: (T_a\phi)(x,y)= \phi(x+a,y)$, we can decompose the
corresponding free Hamiltonian $H_0= -\Delta_D^{\Sigma}$ as
  \begin{equation} 
H_0 = T_a^{-1} \left\{ \bigoplus_{m=-\infty}^{\infty} \tilde
U^{-1} h^{(0)}_m \tilde U \otimes I \right\} T_a
  \end{equation}
with the partial-wave operators
  \begin{equation} \label{partial Ham}
h_m^{(0)} = -{\partial^2\over \partial r^2}\, -{\partial^2\over
\partial y^2}\, + {4m^2-1\over 4r^2}\,, \qquad m\in\Z\,.
  \end{equation}
Their domains are given in a standard way \cite[Sec.~5.7]{BEH};
the only non-trivial part is the radial boundary condition at the
origin. As in \cite[Chap.~I.5]{AGHH} none is imposed in ``higher''
partial waves, $m\ne 0$, because the radial part of (\ref{partial
Ham}) is then limit-point at zero \cite[Appendix to Sec.~X.1]{RS}.
Consequently, this part of the Hamiltonian is trivial from the
point of view of the point interaction. On the other hand, for
$m=0$ we introduce the generalized boundary values,
  \begin{equation} 
\ell_0(\phi)(y):= \lim_{r\to 0}\, \phi(r,y) \sqrt{r}\,, \quad
\ell_1(\phi)(y) := \lim_{r\to 0} r^{-1/2} \bigl\lbrack \phi(r,b)-
\ell_0(\phi)(y) r^{-1/2} \bigr\rbrack\,.
  \end{equation}
The s-wave component of the free Hamiltonian is specified by the
condition $\ell_0(\phi)(y)=0$ (for any $y\in (0,d)$), while the
s-wave component $h_0^{\alpha}$ of $H(\alpha, \vec a)$ is given by
the same differential expression (\ref{partial Ham}) with the
boundary condition at $y=b$ changed to
  \begin{equation} 
\ell_1(\phi)(b) - 4\pi\alpha\ell_0(\phi)(b)= 0\,.
  \end{equation}
It is also clear from this discussion that the eigenfunction of
$H(\alpha, \vec a)$ analyzed in the previous section exhibits a
symmetry, $\psi(R(\varphi)\vec x; \alpha, \vec a)= \psi(\vec x;
\alpha, \vec a)$, where $R(\varphi)$ is the rotation of $\Sigma$
on an angle $\varphi$ around the axis passing through the point
$\vec a$ and perpendicular to $\Sigma$.

Let us return to the scattering problem. As in
\cite[Chap.~I.5]{AGHH} the rotational symmetry means that the
S-matrix part corresponding to partial waves with $m \in
\Z\setminus\{ 0 \} $ is trivial, i.e. the unit operator. In
distinction to \cite[Chap.~I.5]{AGHH}, however, the s-wave
part (we denote it as $S$ again to keep the notation simple) is
still in general a complicated operator because the point
interaction can couple different transverse modes. Its dimension
depends on the number of the ``open channels'', i.e. of the
transverse modes in which the particle of energy $z$ can
propagate; for $d=\pi$ the latter is $[\sqrt{z}]$. Using the
partial-wave operator (\ref{partial Ham}) with $m=0$ it is easy to
see that the function
  \begin{eqnarray}\label{s-wave}
\psi^{(0)}_{\alpha,n}(\vec x;k) &=& J_0(k_n(z) r)\chi_n(y)
\\ \nonumber &\!+\!& {1 \over \alpha \!-\! \xi(\vec a;z)}\, {i \over
2\pi}\, \sum_{j=1}^\infty H_0^{(1)}(k_j(z) r) \sin(jb) \sin(nb)
\chi_j(y)
  \end{eqnarray}
with $r=|x-a|$ satisfies the boundary condition (\ref{bc}) and
$r^{1/2} \psi^{(0)}_{\alpha,n}(\vec x;k)$ is a generalized
eigenfunction of $h_0^{(0)} \otimes \left(-\,{\partial^2 \over
\partial y^2}\right)$ with the eigenvalue $z=k^2$. In the limit $r
\to \infty$ it behaves as
  \begin{eqnarray}
\psi^{(0)}_{\alpha,n}(\vec x;k) &\!\approx\!&  \sqrt{{2 \over \pi
k_n(z) r}}\, \cos ( k_n (z)r - \pi /4) \chi_n(y) \\ \nonumber
&\!+\!& {1 \over \alpha \!-\!\xi(\alpha;z)}\, {i \over 2\pi}\,
\sum_{j=1}^{[\sqrt{z}]} \sqrt{{2 \over \pi k_j(z) r}}\:
e^{i(k_j(z)r -\pi /4)} \sin(nb) \sin(jb) \chi_j (y) \,.
  \end{eqnarray}
To find the matrix elements $S_{nj}$ one has to compare this with
the asymptotics of the outgoing wave in the $j$th transverse mode
expressed by means of the scattering phase shift. For the wave
scattered back into the incident $n$th mode we have
  \begin{eqnarray*}
\sqrt{{2 \over \pi k_n(z) r}} \left[ \cos ( k_n(z)r - \pi /4 ) +
{1 \over \alpha \!-\!\xi(\alpha;z)}\, {i \over 2\pi}\, e^{i (
k_m(z)r -\pi /4)}\, \sin^2(n b) \right] \nonumber \\ = \sqrt{{2
\over \pi k_n(z) r}}\: e^{i \delta_{nn}} \cos (k_n(z)r -\pi /4
+\delta_{nn}(k)) \, , \phantom{AAAAAAAAAAAAA} \nonumber
  \end{eqnarray*}
which yields
$$ S_{nn}(k) = e^{2 i \delta_{nn}(k)} = 1 + {i \over \pi}\,
{\sin^2 (n b) \over \alpha \!-\! \xi(\vec a;z)} \,. $$
In a similar way scattering to the $j$th mode, $n \neq j$,
requires the identification
$$ {1 \over \alpha \!-\!\xi(\vec a;z)}\, {i \over 2\pi}\, e^{i (
k_j(z)r -\pi /4)} \sin(nb) \sin(jb) =  e^{i\delta_{nj}(k)} \cos
(k_n(z)r -\pi /4 +\delta_{nj}(k)) \,; $$
together we get
  \begin{equation} \label{1S}
S_{nj}(k) = e^{2 i \delta_{nj}(k)} = \delta_{nj} + {i \over \pi}\,
{\sin(nb) \sin(jb) \over \alpha \!-\! \xi(\vec a;z)}\,.
  \end{equation}
We have to check that the obtained S-matrix is unitary, i.e.
  \begin{equation}
\sum_{j=1}^{[\sqrt{z}]} S_{nj} \bar S_{sj} = \delta_{ns} \,.
  \end{equation}
Using (\ref{1S}) we can write
  \begin{eqnarray*}
\sum_{j=1}^{[\sqrt{z}]} S_{nj} \bar S_{sj} &\!=\!&
\sum_{j=1}^{[\sqrt{z}]} (S_{nj}-\delta_{nj}) (\bar
S_{sj}-\delta_{js}) + 2 {\rm Re\,} (S_{ns}- \delta_{ns}) +
\delta_{ns} \\ &\!=\!& {2 \over \pi}\, {\sin(n b) \sin(s b) \over
|\alpha \!-\! \xi(\vec a;z)|^2} \left( {1 \over 2\pi}\,
\sum_{j=1}^{[\sqrt{z}]}\sin^2(jb) - {\rm Im\,} \xi(\vec a;z)
\right) + \delta_{ns}\,,
  \end{eqnarray*}
so the desired property follows from (\ref{im part}).

The scattering problem can be also described in another way -- by
means of a scattering operator in $L^2(S^1) \otimes L^2([0,d])$.
Applying (\ref{krein}) to an arbitrary $\phi\in L^2(\Sigma)$ we
see that to any $\psi \in D(H(\alpha, \vec a))$ and a nonreal $z$
there is $\psi_z \in D(H_0)$ such that
  \begin{equation}
\psi (\vec x) = \psi_z (\vec x) + {1 \over \alpha - \xi(\vec
a,z)}\, G_0 (\vec x, \vec a; z) \psi_z (\vec a). \nonumber
  \end{equation}
If we choose, in particular, $\psi_z^\varepsilon (\vec x) = e^{i
k_n(z) \omega x-\varepsilon |x|^2}\chi_n (y)$ with $\omega$ a unit
vector in $\R ^2$, then the corresponding $\psi^\varepsilon (\vec
x) \in D(H(\alpha, \vec a))$ satisfies the equation
  \begin{equation}
\left((H(\alpha,\vec a)\!-\!z)\psi^\varepsilon\right) (\vec x) =
4\varepsilon [1 \!-\! \varepsilon |x|^2 \!+ i k_n \omega x]\,
\psi_z^\varepsilon (\vec x)\,.
  \end{equation}
The r.h.s. converges in the $L^2$ sense as $z$ approaches the real
line and the resulting $\psi^\varepsilon$ belongs to $D(H(\alpha,
\vec a))$ for a fixed $z \in [1,\infty)$. Furthermore, the
pointwise limit exists as $\varepsilon \to 0+$ and equals
  \begin{eqnarray}
\lefteqn{\psi_{\alpha, n} (\vec x; k_n(z) \omega) = e^{i k_n(z)
\omega x} \chi_n (y)}
\\ && +\, {e^{i k_n(z) \omega a} \over \alpha - \xi (\vec a;z)}\, {i
\over 2\pi}\, \sum_{j=1}^\infty H_0^{(1)} (k_j (z) |x-a|) \sin(j
b) \sin(n b) \chi_j (y)\,. \nonumber
  \end{eqnarray}
The function defined by the r.h.s. is locally square integrable,
solves the equation $(H(\alpha,\vec a) \!-\!z)\psi = 0$ and
satisfies the appropriate boundary condition, i.e. it is a
generalized eigenfunction of $H(\alpha,\vec a)$. Let us expand it
into partial waves. We know already the s-wave eigenfunction
(\ref{s-wave}), the remaining ones are trivial, $\psi_n^{(m)}
(\vec x; k_n(z)) = J_m(kr) \chi_n (y)$ if $m\ne 0$. Here again,
this expression describes an eigenfunction of $h^{(0)}_m \otimes
{\partial^2 \over \partial y^2}$ in the Hilbert space (\ref{tensor
Hilb}), multiplying it with $r^{1/2}$ one obtains an eigenfunction
in the Hilbert space (\ref{tensor Hilb2}). Using the known
identity
  \begin{equation}
e^{i k \omega (x-a)} = 2\pi \sum_{m=-\infty}^\infty i^m J_m
(k|x-a|) \overline{Y_m(\omega)} Y_m(\omega_{x-a})
  \end{equation}
where $Y_m$ are the functions introduced above and $\omega_{x-a}:=
{x-a\over|x-a|}$ we get
  \begin{eqnarray*}
\lefteqn{ e^{-i k_n(z) \omega a} \psi_{\alpha, n} (\vec x; k_n(z)
\omega) = \psi^{(0)}_{\alpha, n} (\vec x; k_n(z)) }
\\ && + 2\pi \sum_{0\ne l\in\Z} i^l \psi_n^{(l)} (x-a,y;k_n(z))
\overline{Y_l(\omega)} Y_l (\omega_{x-a})\,.
  \end{eqnarray*}
Components of the on-shell scattering amplitude $\left(
f_{\alpha}(k_n(z), \omega_x, \omega) \right)_{jn}$ are then given
by $\lim_{|x| \to \infty, x / |x| = \omega_x} |x|^{{1 \over 2}}
e^{-i k_j |x|} \left[ \psi_{\alpha, n} (\vec x; k_n(z) \omega) -
e^{i k_n \omega x} \chi_n (y) \right] $, specifically its part
corresponding to the $j$th transverse mode is
  \begin{equation}
\left( f_{\alpha} (k_n(z), \omega_x, \omega) \right)_{jn} =
{e^{\pi i/4} \over \pi \sqrt{2 \pi k_j(z)} }\, {\sin (j b) \sin (n
b) \over \alpha - \xi(\vec a; z)}\, e^{i k_n(z) \omega a - i
k_j(z) \omega_x a}
  \end{equation}
and the on-shell scattering operator $S_{\alpha} (z)$ on $L^2
(S^1) \otimes L^2([0,d])$ has the form
  \begin{equation}
S_\alpha (z) = I + {i \over \pi}\, \sum_{n,j=1}^{[\sqrt{z}]} {\sin
(n b) \sin (j b) \over \alpha - \xi(\vec a; z)}\, \left( e^{-i
k_n(z)(\cdot) a} Y_0\chi_n, \cdot \right) e^{-i k_j(z)(\cdot) a}
Y_0 \chi_j\,.
  \end{equation}
It follows from (\ref{im part}) that the denominator is nonzero in
$[1,\infty)$. However, we have argued above that $\xi(\vec a;
\cdot)$ can be continued analytically to the complex plane where
zeros exist in general. In the weak coupling case there is one
resonance pole of $S_\alpha (z)$ close to the threshold of each
higher transverse mode similarly as in the two-dimensional case
\cite{EGST}.

\setcounter{equation}{0}
\section{A Finite Number of Point Interactions}

\subsection{Boundary conditions}

Consider now a finite number $N$ of point interactions and suppose
that their positions are $\vec a_j = (a_j, b_j)$, where $a_j \in
\R^2$ and $b_j \in (0, \pi)$. For the sake of brevity we denote
$\vec a = (\vec a_1,\ldots,\vec a_N)$ and $\alpha =
(\alpha_1,\ldots,\alpha_N)$. The way to define a point interaction
is the same as above; now we have $N$ independent boundary
conditions
  \begin{equation}\label{bc N}
L_1(\psi,\vec a_j)-4\pi\alpha_j L_0(\psi,\vec a_j) = 0, \qquad
j=1, \ldots, N\,.
  \end{equation}
The Hamiltonian $H(\alpha, \vec a)$ is given again by the formulae
(\ref{one-center perturbation}) and (\ref{one-center domain}) with
the boundary condition (\ref{bc}) replaced by (\ref{bc N}) and
$\vec a$ understood in the sense mentioned above. Any of the point
interactions can be switched off when corresponding coupling
constant $\alpha_j$ is formally put equal to infinity.

\subsection{The resolvent}\label{resol N}

We again use the Krein formula to find the resolvent kernel. Since
the deficiency indices of the operator obtained by restriction of
the free Hamiltonian to the set of functions which vanish at the
vicinity of the points $\vec a$ are equal to $(N,N)$, the r.h.s.
is a rank $N$ operator expressed in terms of vectors from the
corresponding deficiency subspaces,
  \begin{eqnarray} \label{krein N}
\lefteqn{ (H(\alpha,\vec a)-z)^{-1}(\vec x_1,\vec x_2) } \nonumber
\\ && =\, G_0(\vec x_1,\vec x_2 ;z) +\, \sum_{j,k=1}^N
\lambda_{jk}(\alpha,\vec a;z)\,G_0(\vec x_1,\vec a_j ;z)\,
G_0(\vec a_k,\vec x_2 ;z)\,. \phantom{AAA}
  \end{eqnarray}
Applying this to an arbitrary vector of $L^2(\Sigma)$ we get
  \begin{equation} \label{psi decomposition N}
\psi(\vec x) = \psi_0(\vec x)+ \sum_{j,k=1}^N \lambda_{jk}
(\alpha,\vec a;z)\,G_0(\vec x,\vec a_j ;z)\,\psi_0(\vec a_k)
  \end{equation}
with $\psi_0\in D(H_0)$. The generalized boundary values are
  \begin{eqnarray}
L_0(\psi, \vec a_m) &\!=\!& \sum_{j,k=1}^N {\lambda_{jk} \over
4\pi} \, \delta_{jm} \, \psi_0(\vec a_k)\,, \\ L_1(\psi, \vec a_m) &\!=\!& \psi_0
(\vec a_m) + \sum_{j,k=1}^N \lambda_{jk} \delta_{jm} \lim_{|\vec x
- \vec a_m| \to 0} \left( G_0 (\vec x, \vec a_j; z) - {1 \over
4\pi |\vec x - \vec a_m |} \right) \nonumber \\ && +
\sum_{j,k=1}^N \lambda_{jk} (1-\delta_{jm}) G_0 (\vec a_m, \vec
a_j; z) \psi_0 (\vec a_k)\,.
  \end{eqnarray}
The limit contained in the expression of $L_1 (\psi, \vec a_m)$
equals to $\xi (\vec a_m; z)$. After substituting these boundary
values into (\ref{bc N}) we arrive at the conditions
  \begin{equation}
\psi_0 (\vec a_m) + \sum_{j,k=1}^N \lambda_{jk} \left[ \delta_{jm}
(\xi (\vec a_m; z) - \alpha_m) + (1-\delta_{jm}) G_0 (\vec a_m,
\vec a_j; z) \right] \psi_0 (\vec a_k)  =  0\,,
  \end{equation}
which should be satisfied for an arbitrary vector $\psi_0$
belonging to $D(H_0)$, i.e. for any $N$-tuple $\left(\psi_0(\vec
a_1),\dots, \psi_0(\vec a_N) \right)$. This is possible only if
the expressions in the square brackets are up to the sign elements
of the matrix inverse to $\lambda (\alpha, \vec a; z)$, in other
words, if the coefficients are
  \begin{equation}
\lambda (\alpha, \vec a; z)  =  \Lambda (\alpha, \vec a;
z)^{-1}\,,
  \end{equation}
where
  \begin{eqnarray}\label{Lambda}
 \Lambda_{jj} &\!:=\!& \alpha_j - \xi (\vec a_j; z) \\ &\!=\!&\!
 \alpha_j + {1 \over \pi^2} \sum_{n=1}^\infty \ln \sqrt{1- {z
 \over n^2}} \, \sin^2 (n b_j) \,-\, {1 \over 4\pi^2}\! \left[ \gamma
 + \psi \left( {b_j \over \pi} \right) + {\pi \over 2} \cot (b_j)
 \right] \nonumber \\
 \Lambda_{jm} &\!:=\!& - G_0 (\vec a_j, \vec a_m; z) \nonumber
 \\ &\!=\!& - {i \over 2\pi} \sum_{n=1}^\infty H_0^{(1)} \left(
 \sqrt{z-n^2} |a_j-a_m| \right) \, \sin(n b_j) \sin(n b_m)\,, \qquad j
 \neq m\,. \nonumber
  \end{eqnarray}
The Green's function value $G_0 (\vec a_j,\vec a_m; z)$ is finite
for any pair of mutually different vectors $\vec a_j$ and $\vec
a_m$. When the point interactions are arranged vertically, i.e.
$a_j=a_m$, the expression through Hankel's functions is useless
and the corresponding nondiagonal element $\Lambda_{jm}$ can be
alternatively written as
  \begin{eqnarray}\label{Lambda 0}
\Lambda_{jm} &\!=\!& {1 \over \pi^2} \, \sum_{n=1}^{\infty}\, \ln
\sqrt{1-{z \over n^2}} \, \sin(nb_j) \sin(nb_m) \\ && - \xi_2
\left( {b_j+b_m \over 2} \right) + \xi_2 \left( {|b_j-b_m| \over
2} \right)\,. \nonumber
  \end{eqnarray}
To derive this expression we employ the argument analogous to that
leading to the value of the function $\xi(\vec a;z)$ in Section
\ref{resol 1}.

\subsection{The discrete spectrum}\label{spec N}

Since a finite rank operator is both compact and trace class,
the argument presented at the opening of Section~\ref{spec 1}
remains valid. In other words, a finite number of point
interaction changes neither the essential nor the absolutely
continuous spectrum, $\sigma_{ess} (H(\alpha, \vec a)) =
\sigma_{ac} (H(\alpha, \vec a)) = \sigma_{ess} (H_0) = [1,
\infty)$. The singularly continuous spectrum is empty because
the proof given in Sec.~\ref{spec 1} can be used here again.

The discrete spectrum is determined by poles of the resolvent,
which occur if the coefficient matrix $\left(\lambda_{jk}
\right)^{-1}$ becomes singular. This leads to the condition
  \begin{equation}\label{det}
\det \Lambda (\alpha, \vec a; z)  =  0\,.
  \end{equation}
To find the eigenfunctions we use the procedure from
\cite[Sec.~II.1]{AGHH}. Suppose that $H := H(\alpha, \vec a)$
satisfies the equation $H \varphi = z \varphi$  for some $z \in
\R$ and pick an arbitrary $z' \in \varrho (H)$ . Then in
accordance with (\ref{psi decomposition N}) there is a function
$\psi_0 \in D(H_0)$ which makes it possible to write
  \begin{equation}\label{eigen decomposition}
\varphi (\vec x)  =  \psi_0 (\vec x) + \sum_{j=1}^N d_j G_0 (\vec
x, \vec a_j; z')\,,
  \end{equation}
with the coefficients $d_j := \sum_{k=1}^N (\Lambda
(z')^{-1})_{jk} \psi_0 (\vec a_k)$. We also see that
  \begin{equation}
(H_0 - z') \psi_0  =  (H-z') \varphi  =  (z-z') \varphi .
\nonumber
  \end{equation}
Applying the resolvent $(H_0 -z')^{-1}$ to the last identity we
arrive at the expression
  \begin{equation}
\psi_0  =  (z-z') \left[ (H_0 -z')^{-1} \psi_0 + \sum_{j=1}^N d_j
(H_0 -z')^{-1} G_0 (\cdot, \vec a_j; z') \right]\,,
  \end{equation}
which allows us to find the action of $(H_0-z)$ at the vector
$\psi_0$,
  \begin{equation}
(H_0 -z) \psi_0  =  (H_0 -z') \psi_0 - (z-z') \psi_0  =  (z-z')
\sum_{j=1}^N d_j G_0 (\cdot, \vec a_j; z')\,.
  \end{equation}
If $z<1$ , the resolvent $(H_0 -z)^{-1}$  exists and may be
applied to the last relation, giving by means of the first
resolvent identity
  \begin{equation}\label{psi0 sum}
\psi_0  =  \sum_{j=1}^N d_j [G_0 (\cdot, \vec a_j; z) - G_0
(\cdot, \vec a_j; z') ]\,.
  \end{equation}
Substituting this into (\ref{eigen decomposition}) we get an
expression for the eigenfunction,
  \begin{equation}\label{efN}
\varphi (\vec x)  =  \sum_{j=1}^N d_j G_0 (\vec x, \vec a_j; z)\,,
  \end{equation}
where it remains to determine the coefficients. The equation
(\ref{psi0 sum}) taken at the point $\vec x = \vec a_j$ can be
rewritten using the components of matrix $\Lambda$  as
  \begin{equation} \label{psi_0}
\psi_0 (\vec a_j)  =  \sum_{m=1}^N d_m \left( \Lambda (\alpha,
\vec a; z')_{jm} - \Lambda (\alpha, \vec a; z)_{jm} \right)\,.
  \end{equation}
We already know that $d_m = \sum_{k=1}^N (\Lambda (z')^{-1})_{mk}
\psi_0 (\vec a_k)$; taking the inverse we get $\psi_0 (\vec a_j) =
\sum_{m=1}^N \Lambda (z')_{jm} d_m$. In combination with
(\ref{psi_0}) this yields
  \begin{equation}\label{eigenvector}
\sum_{m=1}^N \Lambda (z)_{jm} d_m  =  0,
  \end{equation}
i.e., $d := (d_1, \ldots, d_N)$ has to be an eigenvector of the
matrix $\Lambda (\alpha, \vec a; z)$  corresponding to zero
eigenvalue. The corresponding system of linear equations is
solvable under the condition (\ref{det}), and the sought
eigenfunctions of $H (\alpha, \vec a) $ are given by the formula
(\ref{eigen decomposition}).

It remains to check that the equation (\ref{det}) can have a
solution in $(-\infty,1)$. Let us start with the limiting
situations of strong and weak coupling. We know from (\ref{strong
c asympt}) how $\xi (\vec a_j; z)$  behaves as $-{\sqrt{-z} \over
4\pi}$ as $z \to -\infty$. At the same time, the nondiagonal part
of matrix $\Lambda$ vanishes in the limit in view of the
asymptotics
  \begin{eqnarray}
H_0^{(1)} (\sqrt{z-n^2} |a_j - a_m|) &=& H_0^{(1)} (i \sqrt{n^2-z}
|a_j - a_m|) \nonumber \\ &\approx & \sqrt{{2 \over i\pi
\sqrt{n^2-z}\, |a_j - a_m |}}\: e^{-\sqrt{n^2-z} |a_j - a_m|
-i\pi/4 }  \nonumber
  \end{eqnarray}
for $z\to -\infty$. This argument is not applicable if $a_j=a_m$
for $j \neq m$. Nevertheless, the nondiagonal matrix elements are
up to the sign equal to Green's function values, and thus they are
bounded. In this way we get
  \begin{equation} \label{strong N}
\Lambda (\alpha, \vec a; z)  =  {\sqrt{-z} \over 4\pi}\: I + \OO(1)
\qquad {\rm as} \qquad z \to -\infty
  \end{equation}
with the coefficient $\alpha_j$ included into the error term. On
the other hand,
  \begin{equation}\label{weak N}
\Lambda (\alpha, \vec a; z)  =  {1 \over \pi^2} \ln \sqrt{1-z}\: M
+ \OO(1) \qquad {\rm as} \qquad z \to 1-\,,
  \end{equation}
where $M := \left( \sin b_j \sin b_m \right)_{j,m=1}^N$. This
matrix has zero eigenvalue of multiplicity $N-1$  and the positive
eigenvalue $\sum_{j=1}^N \sin^2 b_j$ corresponding to the
eigenvector $(\sin b_1, \ldots, \sin b_N)$. The latter is more
important, because it means that one eigenvalue of $\Lambda
(\alpha, \vec a; z)$  tends to $ -\infty$  as $z \to 1-$ .
Furthermore, the eigenvalues of $\Lambda (\alpha, \vec a; z)$ are
continuous functions of $z$, so comparing the last claim with
(\ref{strong N}) we find that at least one of the eigenvalues
crosses zero for some $z$, i.e. that $H(\alpha, \vec a)$ has at
least one isolated eigenvalue.

We may ask whether there some of the eigenvalues may be
degenerate. For the sake of brevity we rewrite the matrix
$\Lambda(\alpha, \vec a;z)$ as
  \begin{equation}
\Lambda(\alpha, \vec a;z) = \delta_{jm} (\alpha_j -\xi_j(z)) +
(1-\delta_{jm}) g_{jm}(z), \nonumber
  \end{equation}
where $\xi_j(z) := \xi(\vec a_j;z)$ and $g_{jm}(z) := -G_0 (\vec
a_j, \vec a_m;z)$. Since all $g_{jm}$ are negative, the maximum
possible degeneracy is $N-1$ which means that for a pair of point
interactions the discrete spectrum is always simple. Let us
consider the case $N=3$. Put $z<1$, $\,\vec a_{1,3} =(\pm a, b_1)$
and $\vec a_2 = (0,b_2)$. If $b_2 \to 0$ then the value
$g_{12}=g_{23}$ approaches zero being thus smaller than $g_{13}$
for $b_2$ small enough. On the contrary, if $b_2=b_1$ we obtain
the opposite inequality; this follows from the expression
(\ref{free resolvent}) and from the monotonicity of the Macdonald
function $K_0(u)$ with a positive argument -- see
\cite[9.6.24]{AS}. Hence there exists a $b_2 \in (0,b_1)$ such
that all the non-diagonal elements of the matrix are the same,
$g_{13}=g_{12}=g_{23}$. Choosing $\alpha_j$ which satisfy
$\alpha_j -\xi_j=g_{12}$ we obtain a matrix of rank one, i.e. $z$
is an eigenvalue of multiplicity two.

\subsection{Embedded eigenvalues}

We have shown in the previous chapter that a single point
interaction cannot produce eigenvalues embedded in the continuous
spectrum. This is no longer true if $N\ge 2$ as the following
example shows.
   \begin{example}
{\rm Consider a pair of perturbations with the same $\alpha$
placed at $ \vec a_1 = (0,0,b)$ and $\vec a_2 = (0,0,\pi-b)$. We
can divide the eigenvalue problem into symmetric and antisymmetric
parts with respect to the plane $\{(x,\pi /2)\,: \; x \in \R^2\}
$. We obtain properties of the antisymmetric part by scaling the
one--center problem: substituting $\sigma = {1 \over 2}$ into the
relation (\ref{scale}) we see that the antisymmetric part has a
single eigenvalue which tends to $4$ as $\alpha \to \infty$, hence
it is embedded in the continuous spectrum for $\alpha$ large
enough.}
   \end{example}

Thus we cannot exclude existence of embedded eigenvalues in
general. We can, however, prove a weaker result. In the example
the symmetry was essential, which means in particular that the
eigenfunction is dominated by the second transverse mode. We will
show that in general {\em any eigenvalue $z>1$ cannot contain
contributions from transverse modes with $n \leq [\sqrt{z}]$}.
Suppose that $H(\alpha, \vec a) \varphi = z \varphi$ for $z>1$. We
employ the relation (\ref{eigen decomposition}) and take $\psi_0$
in the form $\psi_0 (\vec x) = \sum_{n=1}^\infty g_n(x) \chi_n
(y)$, where $g_n \in L^2 (\R^2)$. Substituting this $\psi_0$ into
(\ref{eigen decomposition}) and using the fact that $\{ \chi_n \}$
is an orthonormal basis in $L^2 ((0,\pi))$ we get
  \begin{eqnarray*}
\lefteqn{\left[ -{\partial^2 \over \partial x_1^2} -{\partial^2
\over \partial x_2^2} - z +n^2 \right] g_n(x) } \\ && = (z-z')
\sum_{j=1}^N d_j\: {i \over 4} H_0^{(1)} (k_n(z') |x-a_j|) \chi_n
(b_j) \nonumber
  \end{eqnarray*}
for $n=1,2,\ldots$. The Fourier--Plancherel operator transforms
this into
  \begin{equation}\label{fourier}
\left[ |p|^2 - z + n^2 \right] \hat{g}_n (p) = {(z-z') \over 2\pi}
\sum_{j=1}^N d_j\: \chi_n (b_j) {e^{-ipa_j} \over |p|^2 -z'
+n^2}\,.
  \end{equation}
If $g_n \in L^2 (\R^2)$ then $\hat{g}_n$ should also belong to
$L^2 (\R^2)$. It is not possible if $z>n^2$ and the r.h.s. of
(\ref{fourier}) is nonzero at $p_n \omega$, where $p_n =
\sqrt{z-n^2}$ and $\omega$ is a unit vector in $\R^2$. Recall that
the factor $(|p|^2 -z' +n^2)^{-1}$ has no singularity, because $z'
\in \rho(H)$ by assumption.

To avoid the singularity of $\hat{g}_n$ at $|p|=p_n$ we have to
require
  \begin{equation}
\sum_{j=1}^N d_j \chi_n (b_j) e^{-i p_n \omega a_j} = 0
  \end{equation}
for an arbitrary unit vector $\omega$ from $\R^2$. If all the
$a_j$'s are different it follows that $d_j \chi_n (b_j) =0$ for
each $j$. If some of them are the same the condition changes to
$\sum_j d_j \chi_n (b_j) =0$ where $j$ runs through the $a_j$'s
which coincide. In both cases $\hat{g}_n$ is identically zero for
$n^2 <z$.

Consider now an arbitrary $g \in L^2 (\R^2)$ and $n^2 <z$. Using
(\ref{eigen decomposition}) and (\ref{free resolvent}) we arrive
at
  \begin{equation}
(g \chi_n, \varphi) = (\hat{g}, \hat{g}_n) + \sum_{j=1}^N d_j\,
\chi_n (b_j)(g, {i \over 4} H_0^{(1)}(k_n (z') |x-a_j|))\,.
  \end{equation}
The first scalar product is equal to zero because $\hat{g}_n$ is
zero. If all the $a_j$'s are different then $d_j \chi_n (b_j) = 0$
for all $j$ and the whole r.h.s. is equal to zero. If some $a_j$'s
are the same, the condition $\sum_{j=1}^N d_j \chi_n (b_j) = 0$
leads to $(g \chi_n, \varphi) = 0$ again. The conclusion holds
true for all $n=1,\ldots,[\sqrt{z}]$ what we have set out to
prove.

\subsection{The limits of strong and weak coupling}

Of the two extreme situations, consider first the strong coupling.
One can write the matrix $\Lambda (\alpha, \vec a; z)$ in the form
  \begin{equation}
\Lambda(z) = \left( \left( \alpha_j + {\sqrt{-z} \over 4\pi}
\right)\delta_{jk} \right) \left[ I + \left( \left( \alpha_j +
{\sqrt{-z} \over 4\pi} \right)\delta_{jk} \right)^{-1}
\tilde{\Lambda}(z) \right], \nonumber
  \end{equation}
where $ \tilde{\Lambda} (z)$ is the remainder matrix, which is
independent of $\alpha_j$ and has a bounded norm as $z \to
-\infty$. For an arbitrary finite interval $J \subset (-\infty,
1)$ we can always choose $\alpha_j$'s large enough negative that
no eigenvalues are contained in $J$ and the two matrices in the
above product are regular in $J$. It means that the roots of
equation (\ref{point}) come from the region where $\Lambda (z)$ is
dominated by its diagonal part. Then there are exactly N
eigenvalues (including a possible degeneracy), which behave
asymptotically as
  \begin{equation}\label{strong N ev}
\varepsilon_j (\alpha, \vec a) \approx - 16 \pi^2 \alpha_j^2
\qquad {\rm as} \qquad \max_{1\leq j \leq N} \alpha_j \to -
\infty.
  \end{equation}
As we expect the eigenfunctions in the strong limit are strongly
localized and only slightly influenced by the other perturbations.
The eigenfunction localized at $\vec a_j$ has the same form as in
the Section~\ref{spec 1} where we put $\vec a = \vec a_j$.

Consider now on the contrary that all the point interactions are
weak, i.e. that all the $\alpha_j$'s are large positive. Denoting
$A := {\rm diag}(\alpha_1, \ldots, \alpha_N)$ we can write
  \begin{eqnarray}
\Lambda (z) = A + {1 \over \pi^2} \ln \sqrt{1-z}\, M \,+\,
\tilde{\Lambda}(z)
  \end{eqnarray}
where $\tilde{\Lambda} (z)$ is a remainder matrix, which is again
independent of $\alpha$ with its norm bounded,
$\|\tilde\Lambda(z)\| \le C_{\vec a}$ for $z \in (z_0,1)$. In
Section~\ref{spec N} we found that $\Lambda (z)$ was
asymptotically a rank--one operator on $\mathbb{C}^N$. Hence only
one isolated eigenvalue of $H (\alpha, \vec a)$ exists in this
asymptotic situation.

To find the leading term of the asymptotic expansion, we have to
solve the spectral problem for matrix $M - \eta A - \eta
\tilde\Lambda(z)$, where $\eta := -2\pi^2(\ln (1\!-\!z))^{-1}$.
The largest eigenvalue of this matrix satisfies $\mu_N(\eta) \ge
\mu_N(0) - (C_{\vec a}\!+\!\alpha_+)\eta$, where $\alpha_+:= \max
\alpha_j$, while for $j=1,\dots,N\!-\!1$ we have $\mu_j(\eta) \le
(C_{\vec a}\!-\!\alpha_-)\eta$, where $\alpha_-:= \min \alpha_j$.
Since $\mu_N(0)>0$ we see that for $\alpha_-$ large enough the
condition has just one solution for $\eta>0$. One can check
directly that without $\tilde\Lambda(z)$ the condition is
satisfied for $\eta = \sum_j \alpha_j^{-1} \sin^2 b_j$. Thus
$\eta= \OO(\alpha_-^{-1})$ and the eigenvalue expansion follows.
Hence the bound-state energy in weak-coupling case behaves as
  \begin{equation}\label{weak N ev}
\varepsilon (\alpha, \vec a) \approx 1 - \exp\left\{ -2\pi^2
\left( \sum_{j=1}^N {\sin^2 b_j \over \alpha_j} \right)^{-1}
\right\}
  \end{equation}
as $\min_{1\leq j \leq N} \alpha_j \to \infty$. Since the
eigenvector of matrix $M$ corresponding to the nonzero eigenvalue
is $(\sin b_1, \ldots, \sin b_N)$, the asymptotic expression of
the eigenfunction for $\vec x$ from a restricted part of $\Sigma$
is
  \begin{eqnarray}\label{weak N ef}
\lefteqn{\varphi (\vec x) \approx \sin y \left( {\sum_{j=1}^N
\sin^2 b_j \over \sum_{j=1}^N {\sin^2 b_j \over \alpha_j}} \,-\, {1 \over
\pi^2} \sum_{j=1}^N \sin^2 b_j\, \ln |x-a_j| \right) } \nonumber
\\ && +\, {1 \over \pi^2}\, \sum_{n=2}^\infty \sin(n
y)\, \sum_{j=1}^N \sin b_j\, \sin (n b_j)\, K_0 \left(\sqrt{n^2-1}
|x-a_j|\right)\,. \phantom{AAA}
  \end{eqnarray}

Let us summarize the spectral properties of our Hamiltonian in the
$N$-center case derived in the previous three paragraphs.
\begin{theorem}
 Let $H(\alpha, \vec a)$ be defined by (\ref{one-center perturbation})
 and (\ref{one-center domain}) for $d=\pi$, where $\vec a = (\vec a_1,
 \ldots, \vec a_N)$ and $\alpha = (\alpha_1, \ldots, \alpha_N)$
 and the boundary condition (\ref{bc}) is replaced by (\ref{bc N});
 then \\  [1mm]
 (a) $\sigma_{ess}(H(\alpha, \vec a)) = \sigma_{ac}(H(\alpha, \vec a))
 =[1,\infty)$ and $\sigma_{sc}(H(\alpha, \vec a))=\emptyset$.\\ [1mm]
 (b) For any $\alpha\in\R^N$ there is at least one eigenvalue
 $\varepsilon(\alpha,\vec a)$ in $(-\infty,1)$. The maximum
 number of eigenvalues is $N$ with the multiplicity taken into account;
 the maximum multiplicity is $N-1$.
 The corresponding eigenfunction is given by (\ref{efN}), where the
 coefficients $d_m, \; m=1,\ldots,N$, are components of a vector solving
 the equation $\Lambda d=0$ with the matrix $\Lambda$ given by
 (\ref{Lambda}). \\ [1mm]
 (c) In the limit $\min_{1\leq j \leq N} \alpha_j \to \infty$ the
 bound state wave function behaves according to (\ref{weak N ev}) and
 (\ref{weak N ef}). On the other hand, in the strong coupling case
 there are exactly $N$ eigenvalues whose asymptotics is given by
 (\ref{strong N ev}); the corresponding eigenfunctions are strongly
 localized around the points $\vec x = \vec a_j$ and given by
 (\ref{strong 1 ef}) with $\vec a = \vec a_j$ and $\alpha=\alpha_j$. \\ [1mm]
 (d) If an eigenvalue $z\in[1,\infty)$ exists, the corresponding
 eigenvector is orthogonal to the subspace $\bigoplus_{n=1}^{[\sqrt{z}]}
 L^2 (\R^2) \otimes \{ \chi_n \}$.
\end{theorem}


\subsection{Scattering}
Comparing to the one-center case, the Hamiltonian $H(\alpha, \vec
a)$ with a finite number of perturbations loses in general the
invariance with respect to rotations around an axis perpendicular
to $\Sigma$. Hence we cannot employ here the partial wave
decomposition and we turn directly to the ``closed form'' of the
on--shell scattering amplitude and on--shell scattering operator.
By (\ref{psi decomposition N}), to any $\psi \in D(H(\vec
a,\alpha))$ and a nonreal $z$ there exists $\psi_z \in D(H_0)$
such that
  \begin{equation} \label{scatt N}
\psi(\vec x) = \psi_z(\vec x)+ \sum_{j,k=1}^N \lambda_{jk}
(\alpha,\vec a;z)\,G_0(\vec x,\vec a_j ;z)\,\psi_z(\vec a_k).
  \end{equation}
We take again $\psi_0^\varepsilon (\vec x) = e^{i k_n(z) \omega x
-\varepsilon |x|^2} \chi_n (y)$ for $\psi_z$, where $\omega$ is a
unit vector in $\R^2$. Denoting the l.h.s. of (\ref{scatt N}) as
$\psi^\varepsilon$ we have $\psi^\varepsilon \in D(H(\alpha,\vec
a))$ and
  \begin{equation}
\left((H(\alpha,\vec a)\!-\!z)\psi^\varepsilon\right) (\vec x) =
4\varepsilon [1 \!-\! \varepsilon |x|^2 \!+ i k_n \omega x]\,
\psi_z^\varepsilon (\vec x)\,.
  \end{equation}
The r.h.s. converges in $L^2(\R^2)$ as $z$ approaches the real
line and the $\psi^\varepsilon$ belongs to $D(H(\alpha, \vec a))$.
The pointwise limit $\varepsilon \to 0+$ exists and equals
  \begin{equation}
\psi_{\alpha,n}(\vec x; k_n(z) \omega) = e^{i k_n(z) \omega x}
\chi_n (y) + \sum_{j,k=1}^N \lambda_{jk} (\alpha,\vec
a;z)\,G_0(\vec x,\vec a_j ;z)\, e^{i k_n(z) \omega a_k} \chi_n
(b_k).
  \end{equation}
The limiting function is locally square integrable and it thus is
a generalized eigenfunction of $H(\alpha, \vec a)$.

The components $\left( f_{\alpha}(k_n(z), \omega_x, \omega)
\right)_{mn}$ of the on--shell scattering amplitude are then given
by the part of the following expression corresponding to the
outgoing $m$th transverse mode,
$$\lim_{|x| \to \infty, x / |x| = \omega_x} |x|^{1/2} e^{-i k_m
|x|} \left[ \psi_{\alpha, n} (\vec x; k_n(z) \omega) - e^{i k_n
\omega x} \chi_n (y) \right]\,. $$
It yields
  \begin{eqnarray}
\left( f_{\alpha} (k_n(z), \omega_x, \omega) \right)_{mn} &=&
{e^{i\pi/4} \over \pi \sqrt{2 \pi k_m(z)}}\, \sum_{j,k=1}^N e^{-i
k_m(z) \omega_x a_j}\, \lambda_{jk}(\alpha,\vec a;z)\, e^{i k_n(z)
\omega a_k} \nonumber
\\ && \times \:\sin (m b_j) \sin (n b_k)\,,
  \end{eqnarray}
and the on-shell scattering operator $S_{\alpha} (z)$ on $L^2
(S^1) \otimes L^2([0,d])$ is
  \begin{eqnarray} \label{N on-shell}
S_\alpha (z) &=& I + {i \over 2 \pi^2}\, \sum_{k,j=1}^N
\sum_{m,n=1}^{[\sqrt{z}]} \sin (m b_j) \sin (n b_k)\,
\lambda_{jk}(\alpha,\vec a;z)\, \left( e^{-i k_n(z)(\cdot) a_k}
\chi_n, \cdot \right) \nonumber \\ && \times \: e^{-i
k_m(z)(\cdot) a_j} \chi_m\,.
  \end{eqnarray}
As in the one-center case, resonances are determined by the poles
in the meromorphic continuation of the matrix-valued function
$(\lambda_{jk}(\alpha,\vec a;\cdot))$.

\setcounter{equation}{0}
\section{A Layer in Magnetic Field}

\subsection{The free Hamiltonian}

In this section, the layer $\Sigma = \R^2 \times (0,d)$ is placed
into a homogeneous magnetic field $\vec B = (0,0,B)$. As usual the
vector potential generating this field can be chosen in different
ways,
e.g. we can employ the symmetric gauge, $\vec A = {1\over 2}(-B x_2,B x_1,0)$.
We again use the decomposition into transverse modes,
  \begin{eqnarray}\label{decomposition B}
H_0^B &=& \bigoplus_{n=1}^\infty h_n^B \otimes I_n, \\ h_n^B &=&
\left( -i{\partial \over \partial x_1} +{1 \over 2}B x_2 \right)^2
+ \left( -i{\partial \over \partial x_2} -{1\over 2}B x_1 \right)^2
+ \left( {\pi n \over d} \right)^2. \nonumber
  \end{eqnarray}
The first two terms at the r.h.s. denoted as $h^B$ describe a
two-dimensional particle in the perpendicular homogeneous field.
The resolvent kernel of such an operator is well known \cite{DMM}:
  \begin{eqnarray*}
\left( h^B - z \right)^{-1} (x,x') &\!=\!&  {1 \over 4\pi}\, \exp
\left( {i B \over 2}\, (-x_1 x'_2 + x_2 x'_1) - {|B| \over 4}\,
|x-x'|^2 \right) \\ && \times\: \Gamma \left( {|B|-z \over 2 |B|}
\right) U \left( {|B|-z \over 2 |B|}, 1; {|B| \over 2}\, |x-x'|^2
\right)\,,
  \end{eqnarray*}
where $U$ is the irregular confluent hypergeometric function
\cite[13.1.33]{AS}. For the sake of brevity we denote the
exponential term in the above formula as $\Phi^B (x,x')$. The
decomposition (\ref{decomposition B}) then yields the sought
resolvent kernel
  \begin{eqnarray}\label{free resolvent B}
G_0^B (\vec x,\vec x';z) &\!\equiv\!& \left( H_0^B - z
\right)^{-1} (x,y; x',y') \nonumber \\ &\!=\! & {1 \over 2 \pi
d}\, \exp \left( {i B \over 2}\, (-x_1 x'_2 +x_2 x'_1) - {|B|
\over 4}\, |x-x'|^2 \right) \nonumber \\ &\times&
\sum_{n=1}^\infty\: \Gamma \left( {|B|-k_n^2(z) \over 2 |B|}
\right) U \left( {|B|-k_n^2(z) \over 2 |B|}, 1; {|B| \over 2}\,
|x-x'|^2 \right) \nonumber
\\ &\times& \sin \left( {n\pi y \over d} \right) \sin \left( {n\pi y'
\over d} \right).
  \end{eqnarray}

\subsection{The perturbed resolvent}

As in the non-magnetic case we start from a single point
interaction located at the point $\vec a\in \Sigma$ and modify to
the present situation the argument of Sec.~\ref{resol 1}. We
employ the fact that {\em locally} the magnetic field means a
regular perturbation of the Schr\"odinger equation; motivated by
this we {\em define} the one-center Hamiltonian in the same way as
in Chap.~2: it acts as free
  \begin{eqnarray}\label{one-center pert B}
\lefteqn{ (H^B (\alpha, \vec a) \psi)(\vec x)} \\ && =
\left(\left[ \left(-i{\partial \over \partial x_1} +{1\over 2}B
x_2\right)^2 + \left( -i{\partial \over \partial x_2} -{1\over 2}B
x_1 \right)^2 - {\partial^2 \over \partial y^2} \right]
\psi\right) (\vec x) \nonumber
  \end{eqnarray}
for $\vec x\ne \vec a$ with the domain changed to
  \begin{eqnarray}\label{one-center domain B}
D(H^B (\alpha, \vec a)) &\!=\!& \{\, \psi \in L^2 (\Sigma
\setminus \{ \vec a \}) : \; H^B (\alpha, \vec a) \psi \in L^2
(\Sigma \setminus \{ \vec a \}),
\\ && \psi (x,0) = \psi (x,d) =0, \; L_1 (\psi,\vec a) - 4\pi
\alpha L_0 (\psi,\vec a) =0 \,\}\,, \nonumber
  \end{eqnarray}
where $L_0$ and $L_1$ in the last relation are again the
generalized boundary values from \cite[Chap.~I.1]{AGHH}.

\begin{remark} \label{mg sing}
{\rm To justify such a definition we have to check that the
resolvent kernel (\ref{free resolvent B}) has the same singularity
as the non-magnetic expression (\ref{free resolvent}) for $\vec
x'\to \vec x$. We have $G^B_0 (\vec x, \vec x'; z) \approx c|\vec
x'-\vec x|^{-1}$ by \cite[Thm.~III.5.1]{Ber} with a non-zero $c$
independent of $z$. To check that the constant has the needed
value it is sufficient to find operators $H_-$ and $H_{\beta}$
with $\beta$ from some interval $(\beta_0,1)$ whose resolvent
kernels are $\sim(4\pi)^{-1}|\vec x'- \vec x|$ around the
singularity and which satisfy the inequalities
  \begin{equation} \label{mg bracket}
H_- \le H_0^B \le \beta^{-1} H_{\beta}\,,
  \end{equation}
since the last named property implies easily
$$ (H_- -z)^{-1} \geq (H_0^B-z)^{-1} \geq \beta (H_{\beta}-\beta
z)^{-1} $$
for a fixed $z<0$ and $c=(4\pi)^{-1}$ follows by contradiction.
For the lower bound we choose the projection to the layer of the
magnetic Schr\"odinger operator $(-i\vec\nabla-\vec A)^2$ in
$L^2(\R^3)$ obtained by removing the Dirichlet boundaries of
$\Sigma$, because the latter is of the form $h_B \otimes I + I
\otimes (-\partial_y^2)$ and we may apply the bracketing argument
\cite[Sec.~XIII.15]{RS} to the non-magnetic part in the
$y$-direction. Its kernel is known \cite{GD, AGK} to be
\begin{eqnarray}\label{free resol B 3d}
  \widetilde{G}_0^B (\vec x, \vec x'; z) &=& {1 \over 4\pi} \left(
  {|B| \over 2} \right)^{1/2} \Phi^B (x,x') \\ && \times \sum_{l=0}^\infty {\exp
  \left[ - (|B|(2l+1)-z)^{1/2} |y\!-\!y'| \right] \over \left( l+{1
  \over 2} - {z \over 2|B|} \right)^{1/2}}\, L_l \left({B \over 2}
  |x\!-\!x'|^2\right), \nonumber
\end{eqnarray}
where the factor $\Phi^B$ is the same as in the relation
(\ref{free resolvent B}) and $L_l$ are the Laguerre polynomials.
We want to prove that $|\vec x -\vec x'| \widetilde{G}_0^B (\vec
x, \vec x'; z)$ tends to $(4\pi)^{-1}$ in the limit $|\vec x -\vec
x'| \to 0$. Putting $|x-x'|=0$, we can neglect the Laguerre
polynomials and the factor $\Phi^B$; it remains to compute the
simplified sum. For a given $z$ there exists an integer number
$l_0$ such that $|B|(2l_0-1)-z>0$.  Then one could split the
series into two parts: the finite sum over $l=0,\ldots,l_0-1$
which is irrelevant for the singularity and the truncated series
with $l=l_0,\ldots\:$. The latter can be estimated as follows,
\begin{equation}
  I_- \leq {|B| \over 4\pi} \sum_{l=l_0}^\infty {\exp
  \left[ - (|B|(2l+1)-z)^{1/2} \varrho \right] \over \left(|B|(2l+1) -
  z \right)^{1/2}} \leq I_+,
\end{equation}
where
$$
  I_{\pm} := {|B| \over 4\pi} \int_{l_0}^\infty {\exp\left[ -(|B|(2l\mp 1)
  -z)^{1/2} \varrho \right] \over (|B|(2l\mp 1)-z)^{1/2}}\, dl \,=\,
  {1 \over 4\pi\varrho}\: e^{-(|B|(2l_0\mp 1)-z)^{1/2}\varrho}\,.
$$
Hence the resolvent $\widetilde{G}_0^B$ has the needed
singularity.

In he opposite direction we add a Dirichlet boundary at $|x|=R$
cutting thus a finite cylinder of $\Sigma$. It is clearly
sufficient to find a bound of the type (\ref{mg bracket}) for the
``planar'' part of the operator and this task is further reduced
to finding bounds for its partial-wave components,
  \begin{equation} \label{mg disc}
\tilde h_m^B = -\, {\partial^2\over \partial r^2} \, -\,{1\over
r}\, {\partial\over \partial r}\,+\, \left( {m\over r} - {Br\over
2} \right)^2
  \end{equation}
on $L^2(0,R)$ with the boundary condition $\lim_{r\to 0} \phi(r)
(r^{1/2} \ln r)^{-1} =0$
 at the origin \cite[Sec.~I.5]{AGHH} and
Dirichlet at $r=R$. A comparison of the potential terms shows that
$ \tilde h_m^B \le \beta^{-1} \tilde h_m^0 $ holds if $ R < \left(
{2\over |B|} (\beta^{-1/2}\! -1)\right)^{1/2} $, so one can choose
for $H_{\beta}$ the non-magnetic Dirichlet Laplacian in the
cylinder of an arbitrarily small radius. This operator has again
the resolvent kernel with the needed singularity \cite{Titch}.
Notice finally that alternative ways to prove this result can be
found in \cite{DO} or derived by techniques from the classical
theory of partial differential equations \cite[Thm.~20.6]{M}.}
\end{remark}

\begin{remark}
{\rm The previous remark still does not answer the question about
the Green's function singularity fully. To explain why it is the
case, recall that the requirement of symmetry of the magnetic
Hamiltonian yields for any functions $\psi_1,\, \psi_2$ from the
domain by means of the Gauss theorem the condition \cite{DO}
$$ \lim_{r \to 0} \int_{S_{\vec a}} \left( - \bar \psi_1 {\partial
\psi_2 \over \partial r} + \psi_2 {\partial \bar \psi_1 \over
\partial r} + 2 i \bar \psi_1 \psi_2 {\vec A \vec r \over r}
\right)dS \,=\,0\,,$$
where $\vec r=\vec x - \vec a$, $r=|\vec r|$ and the integral is
taken over the surface of the sphere $S_{\vec a}$ with center at
$\vec a$ and radius $r$. It is satisfied if the functions have the
following asymptotic behaviour in the vicinity of the point
interaction:
\begin{equation}\label{mg f sing}
\psi(\vec x) = c_0 {1+\vec A (\vec a) (\vec x - \vec a) \over
|\vec x - \vec a|} +c_1+ \OO(|\vec x -\vec a|).
\end{equation}
This motivates us to change the generalized  boundary value $L_1$
to
\begin{equation}\label{L1 mag}
L_1(\psi,\vec a)\,=\, \lim_{|\vec x -\vec a| \to 0} \left[
\psi(\vec x) - L_0(\psi,\vec a)\, {1+i \vec A (\vec a) (\vec x
-\vec a) \over |\vec x -\vec a|} \right].
\end{equation}
This suggests to use for $\xi(B,\vec a;z)$ the following limit
\begin{equation}
\xi(B,\vec a;z)\,=\, \lim_{\vec x \to\vec a} \left[ G_0^B(\vec
x,\vec a;z) - {1+i \vec A (\vec a) (\vec x -\vec a) \over
4\pi|\vec x -\vec a|} \right],
\end{equation}
which has the disadvantage that it is direction-dependent and can
be altered by a gauge change.

However, it is possible to employ the function $\xi(B,\vec a;z)$
defined by means of the pole singularity alone. By (\ref{free
resolvent B}) the Green function $G_0^B(\vec x,\vec a;z)$ has the
form $\exp(i \vec A(\vec a)(\vec x-\vec a)) F(|\vec x-\vec a|;z)$
in the symmetric gauge. In Remark~\ref{mg sing} we have shown that
$F(|\vec x - \vec a|;z)$ has the following asymptotic behaviour
for small $|\vec x-\vec a|$,
$$F(|\vec x-\vec a|;z) = {1 \over 4\pi |\vec x -\vec a|} +c(z) +
\OO(|\vec x-\vec a|),$$
where $c(z):=\lim_{\varrho\to 0}\left[ F(\varrho;z)- {1\over
4\pi\varrho}\right]$. The function $c$ defined in this way can be
understood as $\xi (B,\vec a;z)$ obtained for the Green's function
without the exponential factor; the corresponding generalized
boundary value $L_1$ does not contain the extra part $\vec A (\vec
a) (\vec x-\vec a)$. In the respective asymptotic formula for
complete Green's function one has to multiply the last expression
by two first terms of Taylor series of the exponential factor,
$$G_0^B(\vec x,\vec a;z) = {1+  i \vec A (\vec a) (\vec x -\vec a)
 \over 4\pi |\vec x -\vec a|} +c(z) +
\OO(|\vec x-\vec a|).$$
Using now the modified boundary value $L_1$ we find that
$\xi(B,\vec a;z)=c(z)$. This justifies the choice of the
generalized boundary values made in the beginning of this section;
one has to keep in mind to ignore the exponential term
$\Phi^B(x,a)$ when computing $\xi(B,\vec a;z)$. This argument
remains also valid for a finite number of point interactions,
hence the functions $\xi(B,\vec a_j;z)$ for $j=1,\ldots,N$
contained in matrix $\Lambda(B,\alpha,\vec a;z)$ in Section~\ref{N
mg} below are computed in the way described here.

}
\end{remark}

After this digression we can return to evaluation of the Green's
function of the operator $H^B (\alpha, \vec a)$. By construction
it is given by the Krein's formula,
  \begin{equation}\label{krein B}
G^B (\vec x, \vec x'; z) = G^B_0 (\vec x, \vec x'; z) \,+\, {G^B_0
(\vec x, \vec a; z) G^B_0 (\vec a, \vec x'; z) \over \alpha
-\xi(B,\vec a;z)}\,,
  \end{equation}
where the regularized Green's function $\xi(B,\vec a;\cdot)$ is,
as we have explained in the above remarks, now given by
  \begin{eqnarray}\label{xi0 B}
\lefteqn{ \xi (B,\vec a;z) = \lim_{|\vec x - \vec a| \to 0} \left[
G^B_0 (\vec x, \vec a;z) - {1 \over 4\pi |\vec x - \vec a|}
\right] } \nonumber
\\  &=& \lim_{\varrho \to 0} \Biggl\lbrack {1 \over 2\pi d}
 \sum_{n=1}^\infty \Gamma \left( {|B| -k_n^2 (z) \over 2|B|}
\right) U \left( {|B| -k_n^2 (z) \over 2|B|},1;{|B| \over 2}
\varrho^2 \right) \nonumber
\\ && \qquad \qquad \qquad \times\: \sin^2 \left( {n\pi b \over d} \right)  - {1
\over 4\pi \varrho}\, \Biggr\rbrack\,.
  \end{eqnarray}
One can check directly the consistency requirement
  \begin{equation} \label{zero B}
\lim_{B \to 0} G^B (\vec x, \vec x'; z) = G (\vec x, \vec x'; z)
  \end{equation}
for fixed $\alpha, \vec a$ and $\vec x\ne\vec x'$, where $G
(\cdot, \cdot; z)$ is the non-magnetic Green's function of
Sec.~\ref{resol 1}. It follows easily from a known relation
\cite[13.3.3]{AS} for the confluent hypergeometric function,
 $$
\lim_{u \to \infty} \Gamma (u) U \left( u,1,{s \over u} \right) =
2 K_0 \left( 2\sqrt{s} \right)\,.
 $$

To make use of the Green's function, we have to evaluate the
r.h.s. of (\ref{xi0 B}). We employ again the same trick and split
a part of the series which can be summed explicitly for a general
$\varrho$ while in the remaining series the limit $\varrho\to 0$
can be interchanged with the sum. Since $U(u,1;\cdot)$ and
$K_0(\cdot)$ have both a logarithmic singularity at zero, we
modify the Ansatz of Sec.~\ref{resol 1} writing $\xi(B,\vec a;z) =
\xi_1 + \xi_2$ with
  \begin{eqnarray*}
\xi_1 &:=& \lim_{\varrho \to 0}\, {1 \over 2 \pi d}
\sum_{n=1}^\infty \left[ \Gamma \left( {|B| -k_n^2 (z) \over 2|B|}
\right) U \left( {|B| -k_n^2 (z) \over 2|B|},1;{|B| \over 2}
\varrho^2 \right) \right.
\\  && \left.
 -\, 2 K_0 \left( \varrho {\pi n
\over d} \right) \right] \sin^2 \left( {n\pi b \over d} \right),
\\ \xi_2 &:=& \lim_{\varrho \to 0} \left[ {1 \over 2 \pi d}
\sum_{n=1}^\infty 2 K_0 \left( \varrho {\pi n \over d} \right)
\sin^2 \left( {n\pi b \over d} \right) - {1 \over 4\pi \varrho}
\right].
  \end{eqnarray*}
The function $\xi_2$ is evaluated as above: in analogy with
(\ref{xi2}) we have
 \begin{equation}
\xi_2 = {\gamma \over 4\pi d}+{1 \over 8\pi d}\left( 2\psi \left(
{b \over d} \right)+\pi\cot \left( {\pi b \over d } \right)
\right)\,.
  \end{equation}
As for the first part $\xi_1$, we employ the small $\varrho$
asymptotics for the confluent hypergeometric and Macdonald
functions,
 \begin{eqnarray*}
U (u,1;s) &\!=\!& -\, {1 \over \Gamma(u)} \left[\, \ln s +\psi(u)
-2\psi(1) \,\right] + \OO (|s \ln s|) \qquad {\rm as} \qquad s \to
0\,,
\\
K_0 (s) &\!=\!& - \left[\, \ln {s \over 2} -\psi(1) \,\right] +\OO
(s^2) \qquad {\rm as} \qquad s \to 0
  \end{eqnarray*}
(see \cite[6.7.13]{BE} and \cite[9.6.13]{AS}). Putting them
together we see that the summand behaves for small $\varrho$ as
$$ \ln  \left((\pi n)^2 \over 2 |B| d^2\right) - \psi \left(
{|B|-z+ \left( {\pi n \over d} \right)^2 \over 2|B|} \right) +\OO
\left( |\varrho^2 \ln \varrho| \right). $$
For large $n$ the digamma function $\psi (s) = \ln(s) -2 /s +
\OO(s^{-2})$, so the above expression can be written for large $n$
as
  \begin{eqnarray*}
\lefteqn{-\ln \left(1+ (|B|-z) \left( {d \over \pi n} \right)^2
\right) + {1 \over 2} {2|B| \over |B| -z + \left( {\pi n \over d}
\right)^2} +\OO(n^{-4})} \\ &&  \phantom{AAA} = z \left( {d \over
\pi n} \right)^2 +\OO (n^{-4})\,. \phantom{AAAAAAAA}
  \end{eqnarray*}
Hence the series in the definition of $\xi_1$ converges uniformly
and the limit can be interchanged with the sum giving the sought
formula for the regularized Green's function,
  \begin{eqnarray}\label{xi B}
\xi (B,\vec a;z) &\!=\!& \xi_1 +\xi_2 \nonumber \\ &=& {1 \over
2\pi d} \sum_{n=1}^\infty \left[\, \ln \left((\pi n)^2 \over 2 |B|
d^2\right) - \psi \left( {|B|-z+ \left( {\pi n \over d} \right)^2
\over 2|B|} \right) \right] \sin^2 \left( {\pi n b \over d}
\right) \nonumber
\\ && +\, {1 \over 4\pi d}\, \left[ \gamma + \psi \left( {b \over d}
\right)+ {\pi \over 2} \cot \left( {\pi b \over d} \right)
\right]\,.
  \end{eqnarray}

  \begin{remark} \label{B scaling}
{\rm The scaling behavior for the family $\Sigma^{\sigma}=\R^2
\times [0, d\sigma], \; \sigma>0$, is similar to that of
Remark~\ref{scaling}, however, one has to scale simultaneously the
magnetic field by
  \begin{equation} \label{B scale}
B^{\sigma}  = \sigma^{-2} B\,.
  \end{equation}
In distinction to the previous sections we shall keep a general
$d$ in the following discussion.}
  \end{remark}

\subsection{Spectral properties}\label{spec B 1}
The essential spectrum of $H^B (\vec a, \alpha)$ remains the same
as that of the free Hamiltonian $H^B_0$ which follows easily from
Weyl's theorem \cite[Thm.~XIII.14]{RS}. The latter is in turn
obtained from the essential spectrum of two--dimensional Landau
Hamiltonian, $\sigma (h^B)= \sigma_{ess} (h^B) = \left\{\, |B| (2
m +1): \, m \in \N_0 \right\} $ -- see, e.g. \cite[Thm.~1]{GHS}.
Using the transverse-mode decomposition (\ref{decomposition B}) we
arrive at
  \begin{equation}
\sigma_{ess} \left( H^B \right) = \sigma_{ess} \left( H_0^B
\right) = \left\{ |B| (2 m +1) + \left( {\pi n \over d} \right)^2:
\; m,n\!-\!1 \in \N_0 \right\}.
  \end{equation}
Furthermore, the general properties of self-adjoint extensions
mentioned in Sec.~\ref{spec 1} imply that there is at most one
eigenvalue in each spectral gap of the unperturbed operator, i.e.
between the two neighboring values from $\sigma_{ess} \left( H^B
\right)$ and in the interval $(-\infty, |B|+ (\pi /d)^2)$; to find
these eigenvalues one has to solve the equation
  \begin{equation}\label{point B}
\xi (B,\vec a;z) = \alpha\,.
  \end{equation}
We have already checked that the terms of the series expressing
$\xi(B,\vec a;z)$ decay as $n^{-2}$ so the series converges with
the exceptions of the points where the $\psi$ function has a
singularity, i.e. for any $z \in \R\setminus\sigma_{ess} \left(
H^B \right)$. It is also obvious that $\xi (B,\vec a;z)$ is real
for any such $z$. We have
  \begin{equation}
{d \over d z} \xi (B,\vec a;z) = {1 \over 4 \pi d |B|}
\sum_{n=1}^\infty \psi' \left( {|B| -z + \left(\pi n \over d
\right)^2 \over 2|B|} \right) \sin^2 \left( {\pi n b \over d}
\right),
  \end{equation}
where $\psi'(s) = {d \over ds} \psi(s)$ is the trigamma function.
For large $n$ it behaves as
$$ \psi' \left( {|B| -z + \left(\pi n \over d \right)^2 \over
2|B|} \right) \,=\, {2|B| \over |B| -z + \left(\pi n \over d
\right)^2} \,+\, \OO(n^{-4})\,, $$
so the series converges for $z \in \R\setminus \sigma_{ess} \left(
H^B \right)$ and we are allowed to interchange the sum with the
derivative. The explicit expression \cite[6.4.10]{AS} for the
digamma derivative,
$$ \psi' (s) = \sum_{j=0}^\infty {1 \over (s+j)^2} \qquad {\rm
for} \quad s \neq 0,-1,-2,\ldots\,, $$
shows that $\xi (B,\vec a;z)$ is monotonously growing with respect
to $z$ in each gap. Using the relation \cite[6.3.16]{AS},
$$ \psi (1+s) = -\gamma + \sum_{j=1}^\infty {s \over j (j+s)},
\qquad s \neq -1,-2,\ldots\,, $$
we find that $\xi$ diverges as $z$ approaches any point of the
essential spectrum behaving in its vicinity as
  \begin{equation}
\xi (B,\vec a;z) = - \, {|B| \over \pi d} \, {\sin^2 \left( {\pi n
b \over d} \right) \over z - |B|(2m+1) -\left( {\pi n \over d}
\right)^2} \, + \, \OO (1)
  \end{equation}
provided the considered value from $\sigma_{ess}(H_0)$ is
``non-degenerate'' in the sense that it can be expressed by means
of a single pair of indices $m,n$.
\begin{figure}[!t]
\begin{center}
\includegraphics[height=10cm, width=\textwidth]{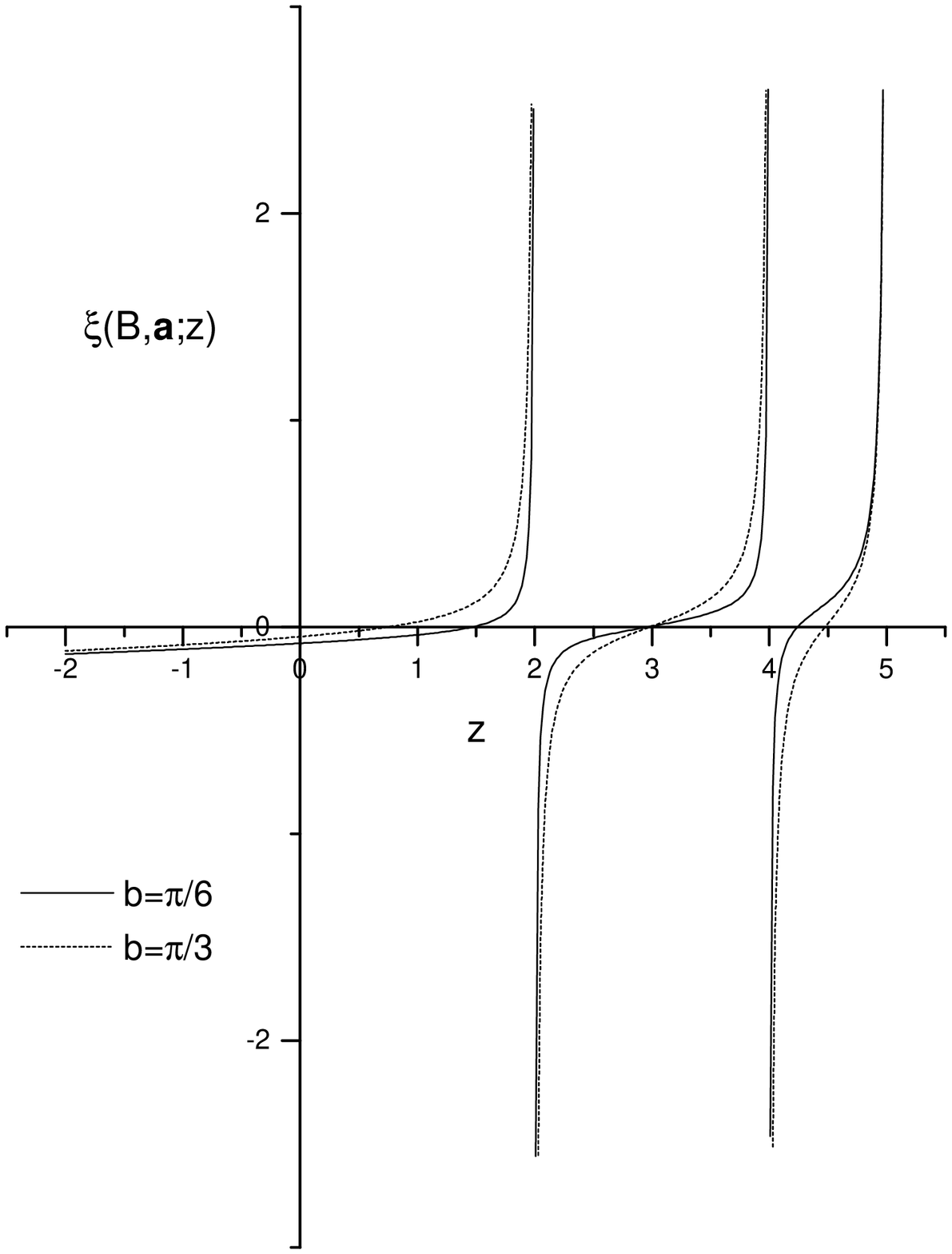}
\end{center}
\vspace{-2cm} \caption{The function $\xi(B,\vec a;\cdot)$ for
$B=1$, $d=\pi$ and $b=\pi/6,\pi/3$.} \label{fig:xi B}
\end{figure}
This is the generical case since the last name property is valid
always if the ratio of the coefficients $|B|$ and $(\pi /d)^2$ is
irrational. If it is rational, then to a given
$z_0\in\sigma_{ess}(H_0)$ there may exist different pairs $m_j,
n_j$ with the index belonging to a family $J(z_0)$ such that
$|B|(2m_j +1) + (\pi n_j / d)^2 = z_0$ for all $j \in J(z_0)$.
Taking this degeneracy into account we have
  \begin{equation}\label{asympt xi B}
\xi (B,\vec a;z) = - \, {|B| \over \pi d} \: {1 \over z-z_0}\,
\sum_{j \in J(z_0)} \sin^2 \left( {\pi n_j b \over d} \right)
\,+\, \OO (1)\,.
  \end{equation}
When $z$ approaches $z_0$ from below, $\xi$ diverges to $\infty$,
while if it approaches $z_0$ from above, $\xi$ goes to $-\infty$.
The formula (\ref{asympt xi B}) also shows that we can disregard
those $j$ for which $\sin (\pi n_j b /d) =0$: as above the system
does not ``feel'' a point perturbation situated at a transverse
eigenfunction node.

To find the behaviour as $z \to -\infty$, recall that the argument
leading to the expression (\ref{xi B}) shows that the latter
differs from the non-magnetic formula (\ref{finalxi}) by the
replacement $z \to z\!-\!|B|$ together with the addition of terms
which remain bounded as $z\to -\infty$. Consequently, the
asymptotics is independent of $B$ and given by the formula
  \begin{eqnarray}\label{strong xi B}
\xi (B,\vec a;z) = - \, {\sqrt{-z} \over 4\pi} \,+\, \OO(1) \qquad
{\rm as} \qquad z \to -\infty.
  \end{eqnarray}

Having discussed the properties of the function $\xi$, we can
apply the conclusions on the equation (\ref{point B}). Since $\xi$
is strictly increasing between every pair of neighbouring
singularities, there is a unique eigenvalue $\varepsilon_r
(B,\alpha,\vec a)$ in each gap for any $\alpha \in \R$ and it
satisfies the inequality
  \begin{equation}
\varepsilon_r (B,\alpha,\vec a) > \varepsilon_r (B,\alpha',\vec a)
\qquad {\rm if} \qquad \alpha > \alpha', \nonumber
  \end{equation}
where the index $r$ is labeling the gaps of $H_0$. The value
$r=0\,$ corresponds to the interval $(-\infty, |B| + (\pi /
d)^2)$, $\,r=1\,$ represents the first finite gap with the left
endpoint $|B| + (\pi / d)^2$, etc.

Since we know the behaviour of $\xi$ around the singularities, we
can write the explicit expression of $\varepsilon_r (B,\alpha,\vec
a)$ in the limits of strong and weak coupling. Suppose that the
point $|B|(2m_j +1) + (\pi n_j / d)^2 = z_0\in\sigma_{ess}(H_0)$
separates the $r$-th and $(r\!+\!1)$-th gap, then using the
implicit-function theorem we get
  \begin{eqnarray} \label{mg asympt}
\varepsilon_r (B,\alpha,\vec a) &=& z_0 - {1 \over \alpha} \,{|B|
\over \pi d} \,\sum_{j \in J(z_0)} \sin^2 \left( {\pi n_j b \over
d} \right) \nonumber \\ &&+ \, \OO (\alpha^{-2}) \qquad {\rm as}
\qquad \alpha \to \infty\,, \\ \varepsilon_{r+1} (B,\alpha,\vec a)
&=& z_0 - \,{1 \over \alpha} \,{|B| \over \pi d} \,\sum_{j \in
J(z_0)} \sin^2 \left( {\pi n_j b \over d} \right) \nonumber
\\ &&+ \,\OO (\alpha^{-2})  \qquad {\rm as} \qquad \alpha \to
-\infty \,. \nonumber
  \end{eqnarray}
We see that the strong and weak limits are similar in the magnetic
case. A different behaviour we find only for the lowest eigenvalue
for which the asymptotic formula (\ref{strong xi B}) gives
  \begin{equation}\label{mg asympt 0}
\varepsilon_0 (B,\alpha,\vec a) = -16 \pi^2 \alpha^2 (1+
\OO(\alpha^{-1})) \qquad {\rm as} \qquad \alpha \to -\infty\,.
  \end{equation}
A finer estimate with the error term replaced by
$\OO(\alpha^{-2})$ can be obtained when the $\xi$ is expressed in
terms of the Hurwitz $\zeta$-function -- see, e.g., \cite{AGK}. As
in the non-magnetic case, the residuum of the resolvent pole in
the formula (\ref{krein B}) yields a (non-normalized)
eigenfunction
  \begin{equation}\label{1 ef B}
\psi (\vec x; \alpha, \vec a) = G^B_0 (\vec x, \vec a;
\varepsilon_r (B,\alpha,\vec a) )
  \end{equation}
corresponding to $\varepsilon_r (B,\alpha,\vec a)$, where $G^B_0$
is the free resolvent (\ref{free resolvent B}).

We are naturally interested in the behaviour of the eigenfunction
in the limit $|\alpha| \to \infty$ for eigenvalues satisfying the
asymptotic relations (\ref{mg asympt}). In the both cases, by
\cite[1.17.11]{BE} the gamma functions in the sum (\ref{free
resolvent B}) corresponding to $n_j, \; j \in J(z_0)\,$ diverges
as
$$ \Gamma (-m_j+\epsilon) \,=\, {(-1)^{m_j} \over m_j !} \:{1
\over \epsilon} \,+\, \OO(1) \qquad {\rm as} \qquad \epsilon \to 0
\,. $$
This makes it possible to write the leading term of the
asymptotics explicitly,
  \begin{eqnarray}\label{mg asympt ef}
\psi (\vec x; \alpha, \vec a) &\!=\!& {\alpha \over \sum_{j \in
J(z_0)} \sin^2 \left( {\pi n_j b \over d} \right)} \:\Phi^B (x,x')
\\ && \!\!\!\!\! \times \sum_{j \in J(z_0)} U \left( -m_j +
{1 \over \alpha} \,{1\over 2 \pi d} \sum_{j \in J(z_0)} \sin^2
\left( {\pi n_j b \over d} \right), 1; {|B| \over 2}\, |x-a|^2
\right) \nonumber \\ && \!\!\!\!\! \times\: {(-1)^{m_j} \over m_j
!} \,\sin \left( {\pi n_j b \over d} \right) \sin \left( {\pi n_j
y \over d} \right) \,+\,\OO(1) \qquad {\rm as} \quad |\alpha| \to
\infty\,. \nonumber
  \end{eqnarray}
 By \cite[13.6.27]{AS} the hypergeometric
function is reduced to a Laguerre polynomial,
 $$
U(-n,1,u) = (-1)^n n! L_n(u)\,, $$
 at positive-integer values of the first index, so  the wave
function can be rewritten as
  \begin{eqnarray}
\psi(\vec x; \alpha,\vec a) &\approx& {\alpha \over \sum_{j \in J(z_0)}
\sin^2 \left( {\pi n_j b \over d} \right)}\, \Phi^B (x,a) \\ && \times \sum_{j
\in J(z_0)}  L_{m_j} \left( {|B| \over 2} |x-a|^2 \right)
\sin \left( {\pi n_j b \over d} \right) \sin \left( {\pi n_j y
\over d} \right). \nonumber
  \end{eqnarray}
In the last formula  we have not specified the error term which
has an extra part coming from the variation of the
hypergeometric function $U(\cdot,1,u)$ around integer values.

We stated that the eigenvalue $\varepsilon_0(B,\alpha,\vec a)$
in the strong coupling limit case has the same behavior as if
there were no magnetic field. The same is true for the
corresponding eigenfunction which is strongly localized,
  \begin{eqnarray}\label{mg asympt 0 ef}
\psi(\vec x; \alpha,\vec a) &=& {1 \over d} \, \Phi^B (x,a)
\sum_{n=1}^\infty \left( 2\pi \sqrt{|B|+ 16\pi^2\alpha^2+ \left(
{\pi n \over d} \right)^2} |x-a| \right)^{-1/2} \\ && \!\!\! \times
e^{-\sqrt{|B|+16\pi^2 \alpha^2+\left( {\pi n \over d} \right)^2}
|x-a|} \sin \left( {\pi n y \over d} \right) \sin \left( {\pi n b
\over d} \right) +\OO(\alpha^{-1}) \nonumber.
  \end{eqnarray}

Let us summarize the results obtained for the layer in the
magnetic field with one point interaction.
\begin{theorem}
 Let $H^B(\alpha,\vec a)$ be defined by (\ref{one-center pert B})
 and (\ref{one-center domain B}); then \\ [1mm]
 (a) $\sigma_{ess}(H^B(\alpha,\vec a))=\left\{ |B|(2m+1)+(\pi
 n/d)^2 \,:\; m,n-1 \in \N_0 \right\}$. \\ [1mm]
 (b) For any $\alpha \in \R$ there exists a single eigenvalue
 $\varepsilon_r (B,\alpha,\vec a)$ between every two neighbouring
 values from the essential spectrum, with the exception of the
 case when the leading term coefficient in (\ref{asympt xi B}) is
 zero and the eigenvalue coincides with the Landau in question for
 a particular value of $\alpha$. Each of these eigenvalues is increasing
 and infinitely differentiable w.r.t. $\alpha$. The corresponding
 eigenfunction is given by (\ref{1 ef B}). \\ [1mm]
 (c) In the both limits $\alpha \to \pm \infty$ the eigenvalue
 behaves similarly according to (\ref{mg asympt}), except for the
 strong coupling limit for the lowest eigenvalue, where the formula
 (\ref{mg asympt 0}) is valid. The eigenfunctions corresponding
 to the eigenvalues (\ref{mg asympt}) and (\ref{mg asympt 0})
 are given by (\ref{mg asympt ef}) and (\ref{mg asympt 0 ef}),
 respectively.
\end{theorem}

\subsection{Finite number of point interactions} \label{N mg}

Next we consider a finite number $N$ of point interactions
supported by points $\vec a_j, \; j=1, \ldots, N$. We define the
Hamiltonian in a way similar to that of Sec.~3, i.e. it will be
given by (\ref{one-center pert B}) and (\ref{one-center domain
B}), where $\vec a$, $\alpha$ are again shorthands for $\vec a =
(\vec a_1, \ldots, \vec a_N)$ and $\alpha = (\alpha_1, \ldots,
\alpha_N)$, and instead of a single boundary condition there is an
$N$-tuple of them,
  \begin{equation}\label{bc N B}
L_1 (\psi, \vec a_j) - 4\pi \alpha_j L_0 (\psi, \vec a_j) =0\,,
\qquad j=1,\ldots,N\,.
  \end{equation}
Accordingly, the Krein's formula reads
  \begin{eqnarray} \label{krein N B}
\lefteqn{ (H^B(\alpha,\vec a)-z)^{-1}(\vec x,\vec x') } \nonumber
\\ && =\, G^B_0(\vec x,\vec x' ;z) \,+\, \sum_{j,k=1}^N
\lambda_{jk}(B,\alpha,\vec a;z)\,G^B_0(\vec x,\vec a_j ;z)\,
G^B_0(\vec a_k,\vec x' ;z)\,. \phantom{AAA}
  \end{eqnarray}
Repeating the argument of Sec.~3.2, we find $ \lambda(B,\alpha,
\vec a;z)^{-1} = \Lambda (B,\alpha,\vec a;z)$ with
  \begin{equation}
\Lambda_{mj} (B,\alpha,\vec a;z) := \delta_{jm} (\alpha_m - \xi
(B,\vec a_m; z) ) - (1 - \delta_{jm}) G^B_0 (\vec a_m, \vec a_j;
z) \,,
  \end{equation}
or more explicitly by means of (\ref{xi B}) and (\ref{free
resolvent B})
  \begin{eqnarray}\label{lambda B d}
\Lambda_{jj} &=& \alpha_j -\, {1 \over 2\pi d}\, \sum_{n=1}^\infty
\left[\, \ln \left((\pi n)^2 \over 2 |B| d^2\right) - \psi \left(
{|B|-k^2_n(z) \over 2|B|} \right) \right] \sin^2 \left( {\pi n b_j
\over d} \right) \nonumber
\\ && -\, {1 \over 4\pi d}\, \left[\, \gamma + \psi \left( {b_j \over d}
\right) +{\pi \over 2} \cot \left( {\pi b_j \over d} \right)
\right]
  \end{eqnarray}
and
  \begin{eqnarray}\label{lambda B n}
\Lambda_{mj} &=&  -\, {1 \over 2\pi d} \,\Phi^B (a_m, a_j)
\,\sum_{n=1}^\infty\: \Gamma \left( {|B|- k^2_n (z) \over 2|B|}
\right) \\ && \times\: U \left( {|B|- k^2_n (z) \over 2|B|},1;{|B|
\over 2}\, |a_m - a_j|^2 \right) \sin \left( {\pi n b_j \over d}
\right) \sin \left( {\pi n b_m \over d} \right)  \nonumber
  \end{eqnarray}
for $j \neq m$. If some perturbations are arranged vertically,
$a_j=a_m$ for $j \neq m$, the last expression can be written as
  \begin{eqnarray}\label{lambda B n 0}
\Lambda_{mj} &=& \, - \,{1 \over 2\pi d} \sum_{n=1}^\infty
\left[\, \ln \left((\pi n)^2 \over 2 |B| d^2\right) - \psi \left(
{|B|-z+ \left( {\pi n \over d} \right)^2 \over 2|B|} \right)
\right] \\ && \times \sin \left( {\pi n b_m \over d} \right) \sin
\left( {\pi n b_j \over d} \right) \nonumber \\ &&  -\, \xi_2
\left({b_j+b_m \over 2} \right) \, + \, \xi_2 \left( {|b_j-b_m|
\over 2} \right) \nonumber
  \end{eqnarray}
as we find by a direct modification of the argument yielding  $\xi
(B,\vec a; z)$.

Consider again a point $z_0\in\sigma_{ess}(H_0)$.
If $z$ approaches this value, the appropriate contributions to the
sums in (\ref{xi B}) and (\ref{free resolvent B}) diverge and the
matrix elements of $\Lambda$ behave in the limit $z\to z_0$ as
  \begin{eqnarray}\label{lambda B lim}
\Lambda_{jj} &=& \alpha_j - \,{|B| \over \pi d} \, {1 \over z_0
-z} \, \sum_{i \in J(z_0)}\, \sin^2 \left( {\pi n_i b_j \over d}
\right) \,+\,\OO(1)\,,
\\ \Lambda_{mj} &=& -\,{|B| \over \pi d} \: \Phi^B (\vec a_m, \vec
a_j)\, {1 \over z_0 -z} \nonumber \\ && \times\: \sum_{i \in
J(z_0)} {(-1)^{m_i} \over m_i!}\: U \left( {|B| + \left( {\pi n_i
\over d} \right)^2 -z \over 2|B|}, 1; {|B| \over 2}\, |a_j -a_m|^2
\right) \nonumber
\\ && \phantom{AA} \times\: \sin \left( {\pi n_i b_j \over d}
\right) \sin \left( {\pi n_i b_m \over d} \right) \,+\,\OO(1) \quad
\mathrm{if} \quad j \neq m\, , \nonumber \\ \Lambda_{mj} &=&
-\,{|B| \over \pi d} \, {1 \over z_0 -z} \sum_{i \in J(z_0)} \,
\sin \left( {\pi n_i b_j \over d} \right) \sin \left( {\pi n_i b_m
\over d} \right) \,+\, \OO(1) \nonumber \\ && \quad \mathrm{if}
\quad j \neq m\ , \; a_j=a_m \, . \nonumber
  \end{eqnarray}
Moreover, the hypergeometric functions reduce to the Laguerre
polynomials in the limit, so
  \begin{eqnarray}
\Lambda_{mj} &\approx& -{|B| \over \pi d} \Phi^B(a_m, a_j) {1
\over z_0 -z} \\ && \times \sum_{i \in J(z_0)} L_{m_i} \left( {|B|
\over 2} |a_m-a_j|^2 \right) \sin \left( {\pi n_i b_j \over d}
\right) \sin \left( {\pi n_i b_m \over d} \right).  \nonumber
  \end{eqnarray}
For large negative energies, $z \to -\infty$, we employ the
asymptotic behavior of $\xi$ given by (\ref{strong xi B}) arriving
thus at
  \begin{equation}
\Lambda_{jj} = \alpha_j \,+\, {\sqrt{-z} \over 4\pi} \,+\,\OO(1)
  \end{equation}
with $B$-dependence being hidden in the error term. The
non-diagonal part of the matrix $\Lambda$ vanishes in the limit of
large negative energy, because by \cite[13.3.3]{AS} one has
  \begin{eqnarray*}
\Lambda_{mj} &\approx& - \,{1 \over 2\pi d} \:\Phi^B (\vec a_m,
\vec a_j) \:\sum_{n=1}^\infty K_0 \left( \sqrt{|B|-k^2_n(z)} |a_m
-a_j |\right) \\ && \phantom{AAAAAAAAAAAA} \times \sin \left( {\pi
n b_j \over d} \right) \sin \left( {\pi n b_m \over d} \right)\,,
  \end{eqnarray*}
which is exponentially decreasing, $K_0 (s) \approx \sqrt{{\pi
\over 2s}}\: e^{-s}\:$ as $s \to \infty\,$. In the case of a
vertical arrangement, $a_j=a_m$ for $j \neq m$, we can again claim
only that the matrix element $\Lambda_{jm}$ is bounded.

Having obtained the coefficient matrix $\Lambda(B,\alpha,\vec
a;z)$ given by (\ref{lambda B d}) and (\ref{lambda B n}) we
proceed with finding the eigenvalues (the essential spectrum is
preserved, of course). In the same way as in Sec.~\ref{spec N} we
check that an eigenvalue $z$ is a solution of the implicit
equation
  \begin{equation}\label{det B}
\det \Lambda(B,\alpha,\vec a;z) \,=\, 0
  \end{equation}
and the corresponding eigenfunction can be written as
  \begin{equation}\label{N ef B}
\varphi(\vec x)=\sum_{j=1}^N d_j G^B_0 (\vec x, \vec a_j;z)\,,
  \end{equation}
where $d:=(d_1,\ldots,d_N)$ is an eigenvector of the matrix
$\Lambda(B,\alpha,\vec a;z)$.

As in the Sec.\ref{spec B 1} we would like to say something about
the number of eigenvalues in each gap. We already know that
this number is limited from above by $N$.
Denoting $\alpha_j = \widetilde{\alpha}_j +\bar{\alpha}$ we can
split the matrix $\Lambda(z,\alpha)$ into two parts
  \begin{equation}
\Lambda(z,\alpha) \,=\, \bar{\alpha}I -M(z,\widetilde{\alpha})\,,
  \end{equation}
where $\widetilde{\alpha} := (\widetilde{\alpha}_1,
\ldots,\widetilde{\alpha}_N)$. The explicit form of the matrix $M$
can be obtained from the relations (\ref{lambda B d}) and
(\ref{lambda B n}) by changing the signs and substituting
$\alpha_j$ by $\widetilde{\alpha}_j$. The same can be done for the
formulae (\ref{lambda B lim}) which express the behavior around
the singularity. In the limit $z \to z_0$ one can neglect the
parameters $\widetilde{\alpha}_j$'s, then it is possible to write
$$ M(z,\widetilde{\alpha}) = {1\over z_0-z}\, {|B| \over \pi d} \,
\bar{M}(z_0)\,+\, \OO(1)\,, $$
where
  \begin{eqnarray}
\bar{M}_{jj}(z_0) &=& \sum_{i \in J(z_0)}\,
\sin^2 \left( {\pi n_i b_j \over d} \right) \nonumber \\
\bar{M}_{mj}(z_0) &=& \Phi_B (\vec a_m,\vec a_j)
\sum_{i \in J(z_0)}\, L_{m_i} \left( {|B| \over 2}|a_m-a_j|^2 \right)
\nonumber \\ && \phantom{AAAAAAAAA} \times
\sin \left( {\pi n_i b_j \over d} \right) \,
\sin \left( {\pi n_i b_m \over d} \right). \nonumber
  \end{eqnarray}
Following \cite{KL} the eigenvalues of $\Lambda(z)$ monotonously
increase between two neighbouring singularities. Then all
eigenvalues of $\bar{M}$ are non-negative numbers. For
$\sum_{j=1}^N \sum_{i\in J(z_0)}\sin^2(\pi n_i b_j/d)\ne 0$ at
least one is positive, hence at least one eigenvalue of the matrix
$M(z,\widetilde{\alpha})$ diverge to $-\infty$ or to $\infty$ as
$z \to z_{0+}$, $z \to z_{0-}$ respectively. Let us summarize the
results:

\begin{theorem}
 Let the operator $H^B(\alpha,\vec a)$ be defined by (\ref{one-center pert B})
 and (\ref{one-center domain B}), where $\vec a = (\vec a_1, \ldots,
 \vec a_N)$ and $\alpha = (\alpha_1, \ldots, \alpha_N)$ and the
 boundary condition is replaced by(\ref{bc N B}); then \\ [1mm]
 (a) $\sigma_{ess}(H^B(\alpha,\vec a))=\left\{ |B|(2m+1)+(\pi
 n/d)^2 \,:\; m,n-1 \in \N_0 \right\}$. \\ [1mm]
 (b) For any $\alpha \in \R^N$ there exists
 at most $N$ eigenvaluee between every two neighbouring
 values from the essential spectrum with the multiplicity taken
 into account.
 The corresponding eigenfunctions are given by (\ref{N ef B}). \\ [1mm]
 (c) In the limits $\min_{1 \leq j \leq N} \alpha_j \to
 \infty$ and $\max_{1\leq j\leq N} \alpha_j \to -\infty$
 at least one eigenvalue converges to each value $z_0$ from the
 essential spectrum from below and from above, respectively, with
 the exception of the case when the leading term coefficients in
 (\ref{lambda B lim}) are zero. In the
 strong limit case there are also $N$ eigenvalues given by
 (\ref{mg asympt 0}) with corresponding eigenfunctions given by
 (\ref{mg asympt 0 ef}), where $\alpha$ and $\vec a$ is replaced
 by $\alpha_j$ and $\vec a_j$ for $j=1 \ldots N$.
\end{theorem}
%
Further results on the number of eigenvalues of the Hamiltonian
$H_0^B$ can be obtained in the same way as in \cite{AG}.

\begin{figure}[!t]
\begin{center}
\includegraphics[height=6cm, width=8cm]{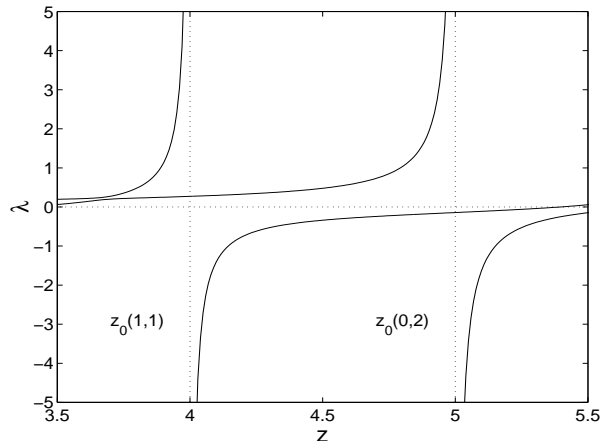}
\end{center}
\vspace{-1cm}
\caption{The behavior of two eigenvalues of the
matrix $M(z,\widetilde{\alpha})$ for $z \in (4,5)$.} \label{fig:gap}
  \end{figure}

  \begin{remark}
{\rm If some of the point interactions are vertically arranged, we
cannot exclude that eigenvalues are absent in a particular gap for
some $\alpha$.

Consider two point interactions placed at $(0,0,1/4 d)$ and
$(0,0,2/3 d)$ with $\widetilde{\alpha}_1=\widetilde{\alpha}_2 =0$.
The numerical calculation for $B=1$ and $d=\pi$ shows that the
eigenvalues of matrix $M(z,\widetilde{\alpha})$ for $z \in
(z_0(1,1),z_0(0,2))$ cover whole $\R$ except one gap, see Fig.
~\ref{fig:gap}. The symbol $z_0(m,n)$ represents the Landau level
$z_0=|B|(2m+1)+(\pi n /d)^2$. Hence for $\bar{\alpha}$ from this
gap the matrix $\Lambda$ has no eigenvalue in the interval
$(z_0(1,1),z_0(0,2))=(4,5)$. }
  \end{remark}


\setcounter{equation}{0}
\section{Conclusions}

We have analyzed here spectral and scattering properties of a
hard-wall layer with a finite number of point interactions. The
results offer one more illustration of efficiency of Krein's
formula which allows to reduce the task in fact to an algebraic
problem. There are other interesting question related to systems
with finitely many perturbations such as relations between the
perturbation configurations and the spectra including the nodal
structure, etc., positions of resonances including those coming
from perturbation of the embedded eigenvalues, and so on. To keep
this paper within reasonable limits, however, we postpone these
questions to a sequel.

The same applies to systems with an infinite number of point
obstacles which offer a wider variety of spectral types. Let us
briefly mention several problems which we regard as worth of
attention. One of them concerns the number of gaps in periodic
systems. A periodic layer of point interactions in $\R^3$ has at
most one gap \cite[Sec.~III.1]{AGHH}. On the other hand it is
known that the presence of boundaries can enhance the number of
gaps in the two-dimensional case \cite{EGST}; a similar effect is
expected in dimension three: for a thin enough layer there will be
many open gaps between the first and the second transverse
threshold. A more difficult question concerns the validity of the
Bohr-Sommerfeld conjecture in such systems.

Even more interesting are spectral properties of periodically
perturbed layers in presence of a magnetic field. It is well known
that a combination of a square lattice of point interactions and a
homogeneous magnetic field leads to a very rich spectrum whose
properties depend substantially on the number-theoretical
properties of the ratio between the lattice spacing and the field
intensity (which determines the cyclotronic radius) -- see
\cite[Sec.~III.2.5]{AGHH} or \cite{Sh}. Putting such a system into
a layer brings a third parameter (the layer width $d$) which will
determine how ``thickly'' the transverse-mode component are
overlayed in the spectrum.

The same applies to edge-type states. It was shown recently that
an equidistant array of point interaction in combination with a
homogeneous magnetic field can produce bands of absolutely
continuous spectrum away of the Landau levels \cite{EJK}. One is
naturally interested how the spectrum will change if the array is
confined between a pair of hard walls. Other problems concern
aperiodic perturbations, external electric field, spin effects,
etc.


\subsection*{Acknowledgments}

Useful remarks of V.~Geyler and P.~\v{S}eba are gratefully
acknowledged. The research was partially supported by the GAAS
grant \# 1048801.

\bibliographystyle{plain}

\end{document}